\documentstyle[12pt]{article}

\newcommand{\Od}{{\cal O}}

\begin{document}
\input epsf \renewcommand{\topfraction}{0.8}
\pagestyle{empty} \begin{flushright}
{LBNL-51861}
\end{flushright}
\vspace*{5mm}
\begin{center}
\Large{\bf Limits on the brane fluctuations mass and on the brane 
tension
scale from electron-positron colliders}
\\ \vspace*{2cm}
\large{\bf  J. Alcaraz$^{(a)}$, J. A. R. Cembranos$^{(b)}$, 
A. Dobado $^{(b,c)}$  
\\ and A. L.
Maroto $^{(b)}$}\\
\vspace{0.2cm} \normalsize
{\it (a) Divisi\'on de  F\'{\i}sica de Part\'{\i}culas,
CIEMAT. 28940 Madrid, Spain}
 \\ \vspace{0.2cm} \normalsize
{\it (b) Departamento de  F\'{\i}sica Te\'orica,\\
 Universidad Complutense de
  Madrid, 28040 Madrid, Spain}\\
 \vspace{0.2cm} \normalsize 
{\it (c) Theory Group, Lawrence Berkeley National Laboratory \\
 Berkeley, CA 94720, USA}\\   
\vspace*{0.6cm} {\bf ABSTRACT} \\
\end{center}
In the context of the  brane-world scenarios with compactified 
large extra
dimensions, we study the production of the possible  
massive brane oscillations
(branons) in electron-positron colliders. We compute their 
contribution to the
electroweak gauge bosons decay width and to the single-photon 
and single-Z processes. With LEP-I results and assuming non 
observation at LEP-II we present exclusion 
plots for the
brane tension $\tau = f^4$ and the branon mass $M$. Prospects 
for the next
generation of electron-positron colliders are also considered.

\vspace*{5mm}

\noindent
\begin{flushleft} PACS: 11.25Mj, 11.10Lm, 11.15Ex \\
\end{flushleft}
\newpage
\setcounter{page}{1} \pagestyle{plain}
\textheight 20 true cm
\section{Introduction}

In the last years a lot of attention has been paid to the so called 
brane
world or ADD scenarios \cite{ADD} where the Standard Model 
particles are
confined to live in  the world brane and only gravitons are free 
to move along
the $D$ dimensional bulk space (see \cite{rev} for recent reviews). 
The
phenomenological consequences of these real or virtual gravitons, 
described in
terms of their corresponding Kaluza-Klein (KK) towers, 
in colliders or in 
astrophysical and cosmological scenarios,  have been the object 
of many recent
works (see \cite{HS} and references therein). However, in 
addition to the
gravitons and the Standard Model (SM) particles, one has to 
consider in
principle the possibility of having also brane oscillations (branons). 
In fact, 
it has been shown that these branons or, equivalently, the brane recoil,
give rise to an exponential suppression of the couplings of the 
SM particles
and the higher KK modes \cite{GB} in such a way that, in the 
$f\ll M_F$ regime
(where $\tau=f^4$ is the brane tension and $M_F$ is the $D$ dimensional
fundamental scale of gravity), the most important  modes at 
low energies are the SM
particles and the branons. Moreover branons can play a role in the 
solution of
some problems appearing when the flexibility of the brane is not taken into
account such as divergent virtual contributions from the KK tower or 
non-unitary graviton production cross-sections. 

The effective action for  the SM fields on the brane was obtained in
\cite{Sundrum}. The introduction of brane fluctuations was done in
\cite{DoMa}, where the interaction between branons and the SM particles, and
the branons self interactions, were obtained including also the possibility of
having non-vanishing branon masses due to the non-factorization of the
extra dimension space.

The effects of brane recoil on real graviton production have been
studied  for example in \cite{MU}, and on virtual gravitons and gauge
bosons in \cite{Contino,BT}. Some constraints from astrophysics on the brane tension
were considered in  \cite{Kugo} and the direct production of branons in
colliders was discussed in \cite{CrSt}, for massless branons in both cases.

In this work we are interested in the production rates of massive 
branons in
electron-positron colliders, in order to get bounds on the
brane tension and the branon mass in the above mentioned 
scenario where the
brane tension scale $f$ is much smaller that the $D$ dimensional 
fundamental
scale of gravity $M_F$. The paper is organized as
follows: in Sec. II we define our set up and give 
the effective
action for massive branons. In Sec. III we obtain their couplings 
to the SM
particles and in Sec. IV we discuss the kind of bounds which
is possible to
set on the brane parameters, namely the number of branons $n$, 
their mass $M$
and the brane tension scale $f$. Sec. V is devoted to the 
analysis of the Z
invisible width and Sec. VI to the $W$ decay. Direct searches 
based on single-photon and single-Z processes are considered 
in Sec. VII, and extended to future linear colliders in Sec VIII. 
Sec. IX offers the summary and the main conclusions of this work. 
In addition, the
relevant Feynman rules are shown in App. A. In App. B
the probability amplitudes are calculated for the relevant
processes and in App. C, 
it is possible to
find some two-body phase-space exact integrals 
which were used in this work. Finally, App. D contains some
explicit expressions for the cross sections.

\section{The effective action for massive brane fluctuations}

We will consider a single-brane model in large extra dimensions. In
such model, our four-dimensional space-time $M_4$ is embedded in a
$D$-dimensional bulk space which, for simplicity, we will assume to
be of the form $M_D=M_4\times B$. The $B$ space is a given
N-dimensional compact manifold, so that $D=4+N$. The brane lies
along $M_4$ and we neglect its contribution to the bulk
gravitational field. The coordinates parametrizing the points in
$M_D$ will be denoted by $(x^{\mu},y^m)$, where the different
indices run as $\mu=0,1,2,3$ and $m=1,2,...,N$. The bulk space
$M_D$ is endowed with a metric tensor which we will denote by
$G_{MN}$, with signature $(+,-,-...-,-)$. For simplicity, we will
consider the following ansatz:
\begin{eqnarray}
 G_{MN}&=&
\left(
\begin{array}{cccc}
\tilde g_{\mu\nu}(x)&0\\ 0&-\tilde g'_{mn}(y)
\end{array}\right).
\label{bulkmetric}
\end{eqnarray}

The position of the brane in the bulk can be parametrized as
$Y^M=(x^\mu, Y^m(x))$, with $M=0,\dots, 3+N$ and
where we have chosen the bulk coordinates
so that the first four are identified with the space-time brane
coordinates $x^\mu$. We assume the brane to be created at a
certain point in $B$, i.e. $Y^m(x)=Y^m_0$ which corresponds to its
ground state.

The induced metric on the brane in such state is given by the
four-dimensional components of the bulk space metric, i.e.
$g_{\mu\nu}=\tilde g_{\mu\nu}=G_{\mu\nu}$. However, when brane
excitations  are present, the induced metric is given by
\begin{eqnarray}
g_{\mu\nu}=\partial_\mu Y^M\partial_\nu Y^N G_{MN}(x,Y(x)) =\tilde
g_{\mu\nu}(x,Y(x))-\partial_{\mu}Y^m\partial_{\nu}Y^n\tilde
g'_{mn}(Y(x)).
\end{eqnarray}

Since the mechanism responsible for the creation of the brane is
in principle unknown, we will assume that the brane dynamics can
be described by an effective action. Thus, we will consider the
most general expression which is invariant under
reparametrizations of the brane coordinates. Following the
philosophy of the effective Lagrangians technique, we will also
organize the action as a series in the number of the derivatives
of the induced metric over a dimensional constant, which can be
identified with the brane tension scale $f$. From this point of
view, to the lowest order in derivatives we find:
\begin{equation}
S_B=\int_{M_4}d^4x\sqrt{g}\left( -f^4 + \dots\right),
\label{Nambu4}
\end{equation}
where $d^4x\sqrt{g}$ is the volume element of the brane. Notice
that this lowest order term is the usual Dirac-Nambu-Goto action.
However, in certain circumstances the
effect of the higher order terms must be taken into account 
\cite{BSky}.

In the absence of the 3-brane, the  metric (\ref{bulkmetric})
possesses an isometry group which we will assume to be of the form
$G(M_D)=G(M_4)\times G(B)$. If the brane ground state is
$Y^m(x)=Y^m_0$, the presence of the brane will break spontaneously
all the $B$ isometries, except those that leave the point $Y_0$
unchanged. In other words the group $G(B)$ is spontaneously broken
down to $H(Y_0)$, where $H(Y_0)$ denotes the isotropy group (or
little group) of the point $Y_0$. Therefore, we can introduce the
coset space $K=G(M_D)/(G(M_4)\times H(Y_0))=G(B)/H(Y_0)$.

Let $H_i$ be the $H(Y_0)$ generators ($i=1,2,...\; h$),
$X_\alpha$ ($\alpha=1,2,...\;k=\mbox{dim}\; G(B)-\mbox{dim}\;H$)
the broken generators, and $T=(H,X)$ the complete set of
generators of $G(B)$. A similar separation can be done with the
Killing fields. We will denote $\xi_i$ those associated to the
$H_i$ generators, $\xi_\alpha$ those corresponding to $X_\alpha$
and by $\xi_a(y)$
 the complete set of Killing
vectors on $B$. As shown in \cite{DoMa},
the excitations of the brane along the directions
of the broken generators of $B$ correspond to the zero modes and
they are parametrized by the Goldstone bosons (branon fields)
$\pi^\alpha(x)$, which
can be understood as coordinates on the coset manifold $K$. When 
the $B$ space is homogeneous then the coset $K$ is isomorphic to $B$ and
the isometries are just translations. In this case the branon fields
($\pi$) can 
 be identified with properly chosen coordinates in the extra
space ($Y$).  
 
According to the previous discussion, we can write the induced
metric on the brane in terms of branon fields as:
\begin{equation}
g_{\mu\nu}=\tilde g_{\mu\nu}(x)- \tilde
g'_{mn}\frac{\partial Y^m}{\partial\pi^\alpha}\frac{\partial
Y^n}{\partial\pi^\beta}\partial_{\mu}\pi^\alpha
\partial_{\nu}\pi^\beta.
\end{equation}

Introducing the tensor $h_{\alpha\beta}(\pi)$  as
\begin{equation}
h_{\alpha\beta}(\pi)=f^4 \tilde g'_{mn}(Y(\pi))\frac{\partial
Y^m}{\partial\pi^\alpha}\frac{\partial Y^n}{\partial\pi^\beta},
\end{equation}
we have
\begin{eqnarray}
g_{\mu\nu}&=&\tilde
g_{\mu\nu}(x)-\frac{1}{f^4}h_{\alpha\beta}(\pi)\partial_{\mu}\pi^\alpha
\partial_{\nu}\pi^\beta.
\label{induced}
\end{eqnarray}
Branons are massless only if the isometry pattern introduced
before is exact. However, in
general, the symmetry is only approximately realized and
branons will acquire  mass. In order to show this mechanism
explicitly, let us perturb the four-dimensional components of the
background metric and let $\tilde g_{\mu\nu}$ depend, not
only on the $x$ coordinates, but also on the $y$ ones:
\begin{eqnarray}
 G_{MN}&=&
\left(
\begin{array}{cccc}
\tilde g_{\mu\nu}(x,y)&0\\ 0&-\tilde g'_{mn}(y)
\end{array}\right).
\end{eqnarray}
This has to be done in such a way
that the $G(B)$ piece of the full isometry group is explicitly
broken. Notice that the breaking of the $G(B)$ group by perturbing
only the internal
metric $\tilde g_{mn}'(y)$  does not lead to a mass term for the
branons.

In order to calculate the branon mass matrix, we need to know
first the  ground state around which the brane is fluctuating.
With that purpose, we will
consider for simplicity the lowest-order action, given by:
\begin{eqnarray}
S_{eff}^{(0)}[\pi]&=&\int_{M_4}d^4x {\cal L}^{(0)}= -f^4
\int_{M_4}d^4x\sqrt{\tilde g(x,Y(x))}
\end{eqnarray}
which will have an extremum provided
\begin{eqnarray}
\delta S_{eff}^{(0)}[\pi]=0\Rightarrow \delta\sqrt{\tilde
g}=\frac{1}{2}\sqrt{\tilde g}\tilde g^{\mu\nu}\delta \tilde
g_{\mu\nu}=0\Rightarrow \tilde g^{\mu\nu}\partial_{m} \tilde
g_{\mu\nu}=0, \forall   y^m.
\label{extremum}
\end{eqnarray}
This is a set of equations whose solution $Y^m_0(x)$ determines
the shape of the brane in its ground state for a given background
metric $\tilde g_{\mu\nu}$.
In addition, the  condition for the energy to be minimum requires:
\begin{eqnarray}
\left.\frac{\delta^2 {\cal L}^{(0)}}{\delta Y^m\delta
Y^n}\right|_{Y=Y_0}&=&\frac{-f^4}{4}\sqrt{\tilde g}\tilde
g^{\mu\nu}(\partial_{n}\partial_{m} \tilde g_{\mu\nu}-2\tilde
g^{\rho\sigma}\partial_{n} \tilde g_{\nu\sigma}\partial_{m} \tilde
g_{\mu\rho})<0. \label{condsy}
\end{eqnarray}
i.e. the eigenvalues of the above matrix should be negative. This
implies that the static Lagrangian should have a maximum.

In order to obtain the explicit expression of the branon
mass matrix, we expand  $\tilde g_{\mu\nu}(x,y)$ around
$y^m=Y^m_0$ in terms of the $\pi^\alpha$ fields:
\begin{eqnarray}
\tilde g_{\mu\nu}(x,y)&=&\tilde g_{\mu\nu}(x,Y_0)+\partial_m
\tilde g_{\mu\nu}(x,Y_0)(Y^m-Y^m_0)
\label{indbran} \\
&+&\frac{1}{2}\partial_m\partial_n\tilde
g_{\mu\nu}(x,Y_0)(Y^m-Y^m_0)(Y^n-Y^n_0)+...\nonumber \\ &=& \tilde
g_{\mu\nu}(x,Y_0)
+\frac{1}{f}V_{\alpha\mu\nu}^{(1)}\pi^\alpha
+\frac{1}{f^2}V_{\alpha\beta\mu\nu}^{(2)}\pi^\alpha\pi^\beta
+\frac{1}{f^3}V_{\alpha\beta\gamma\mu\nu}^{(3)}\pi^\alpha\pi^\beta\pi^\gamma
+\dots\nonumber
\end{eqnarray}
%

The linear term in branon fields is written as: 
\begin{eqnarray}
\frac{1}{f}V_{\alpha\mu\nu}^{(1)}=\left.\partial_m\tilde
g_{\mu\nu}(x,y)\right\vert_{y=Y_0}\frac{\xi^m_\alpha}{kf^2},
\end{eqnarray}
while the
quadratic term takes the general form:
\begin{eqnarray}
\frac{1}{f^2}V_{\alpha\beta\mu\nu}^{(2)}=\frac{1}{2}\left.\partial_m 
\tilde
g_{\mu\nu}(x,y)\right\vert_{y=Y_0}\left.\frac{\partial^2
Y^m}{\partial \pi^\alpha\partial \pi^\beta}\right\vert_{\pi=0}
+
\frac{1}{2}\left.\partial_m\partial_n\tilde
g_{\mu\nu}(x,y)\right\vert_{y=Y_0}\frac{\xi^m_\alpha\xi^n_\beta}{k^2f^4}
\end{eqnarray}
and we have not written the explicit expression for
$V_{\alpha\beta\gamma\mu\nu}^{(3)}$ since it will play no role in
the present  work. Here, we have used the fact that the action of an
element of $G(B)$ on $B$ will map $Y_0$ into
some other point with coordinates:
\begin{equation}
Y^m(x)=Y^m(Y_0,\pi^\alpha(x))=Y^m_0+\frac{1}{k
f^2}\xi^m_\alpha(Y_0)\pi^\alpha(x)+\Od(\pi^2)
\end{equation}
where we have set the normalization of the branon
and  Killing fields  with
$k^2=16\pi /M_P^2$ being $M_P$ the four-dimensional Planck mass.
Substituting the above expression back in
(\ref{induced}), we get the expansion of the induced metric in
branon fields:
\begin{equation}
g_{\mu\nu}=
\tilde g_{\mu\nu}(x,Y_0)-\frac{1}{f^4}\delta_{\alpha\beta}\partial_{\mu}\pi^\alpha
\partial_{\nu}\pi^\beta
+\frac{1}{f}V_{\alpha\mu\nu}^{(1)}\pi^\alpha
+\frac{1}{f^2}V_{\alpha\beta\mu\nu}^{(2)}\pi^\alpha\pi^\beta
+\Od(\pi^3).
\end{equation}

We have also used the fact that since $\pi^\alpha$ must be 
properly normalized scalar fields, the $Y^m$ coordinates
should be normal and geodesic in a neighborhood of
$Y^m_0$ and, in particular, they cannot be angular coordinates. 
This
implies that we can write 
$h_{\alpha\beta}(\pi=0)=\delta_{\alpha\beta}$.

Assuming for concreteness that, in the ground state, the
four-dimensional background metric is flat, i.e. $\tilde
g_{\mu\nu}(x,Y_0)=\eta_{\mu\nu}$, the appearance of the
$V_{\alpha_1\alpha_2...\alpha_i\mu\nu}^{(i)}$ tensors
in (\ref{indbran}) could break Lorentz invariance,
unless they factor out as
$V_{\alpha_1\alpha_2...\alpha_i\mu\nu}^{(i)}=
M_{\alpha_1\alpha_2...\alpha_i}^{(i)}\eta_{\mu\nu}/(4f^2)$.
With this assumption, the linear term
$V_{\alpha\mu\nu}^{(1)}$  vanishes identically
due to the condition of minimum for the brane energy (\ref{extremum}), 
and the $M_{\alpha\beta}^{(2)}$ coefficient in the quadratic term
can be identified with the branon mass matrix.
Thus, in general
and for   the square
root of the metric determinant we find:
\begin{equation}
\sqrt{g}=1-\frac{1}{2f^4}\eta^{\mu\nu}\delta_{\alpha\beta}
\partial_{\mu}\pi^\alpha\partial_{\nu}\pi^\beta
+\frac{1}{2f^4}M_{\alpha\beta}^{(2)}\pi^\alpha\pi^\beta
+... \label{det}
\end{equation}
Notice that this expression requires that both $\partial \pi/f^2$
and $M^2\pi^2/f^4$ are small. This includes different types of
expansions, such as low-energy expansions with small branon masses
compared to $f$, or low-energy expansions with possible large
masses and small $\pi/f$ factors.

The different terms in the effective action can be organized according 
to the number of branon fields:
\begin{equation}
S_{eff}[\pi]=S_{eff}^{(0)}[\pi]+ S_{eff}^{(2)}[\pi]+ ...
\end{equation}
where the zeroth order term is just a constant. The free action
contains the terms with two branons:
\begin{eqnarray}
 S_{eff}^{(2)}[\pi]=\frac{1}{2}\int_{M_4}d^4x
(\delta_{\alpha\beta}\partial_{\mu}\pi^\alpha\partial^{\mu}\pi^\beta
-M^2_{\alpha\beta}\pi^\alpha\pi^\beta).
\end{eqnarray}

\vspace{.5cm}

\section{Couplings to the Standard Model fields}

As we have shown in the previous sections, the induced metric on
the brane depends on both the four-dimensional bulk metric
components $\tilde g_{\mu\nu}$ and the branon fields $\pi^\alpha$.
As noticed in \cite{Contino}, in the limit in which gravity
decouples $M_F\rightarrow \infty$, the branon fields still
survive. This implies that their effects can be studied
independently of gravity. With that purpose, in the following, we
will consider their couplings to the Standard Model  fields in
the absence of  gravitational background field i.e. 
$\tilde g_{\mu\nu}=\eta_{\mu\nu}$. In order
to obtain the general couplings, one can proceed as in
\cite{DoMa}, where the action on the brane, which is basically the
SM action defined on a curved space-time $M_4$, is expanded in
branon fields through the induced metric. For example, the
complete action, including terms with two branons, 
contains the SM
term, the kinetic term for the branons and the interaction 
terms
between the SM particles and the branons:

\begin{eqnarray}
S_B&=& \int_{M_4}d^4x\sqrt{g}[-f^4+ {\cal L}_{SM}(g_{\mu\nu})]
\nonumber\\
&=&\int_{M_4}d^4x\left[-f^4+ {\cal L}_{SM}(
\eta_{\mu\nu})  +
\frac{1}{2}\eta^{\mu\nu}\delta_{\alpha\beta}\partial_{\mu}\pi^\alpha
\partial_{\nu}\pi^\beta-\frac{1}{2}M^2_{\alpha\beta}\pi^\alpha\pi^\beta\right.
\nonumber\\
&+& \left.\frac{1}{8f^4}(4\delta_{\alpha\beta}\partial_{\mu}\pi^\alpha
\partial_{\nu}\pi^\beta-M^2_{\alpha\beta}\pi^\alpha\pi^\beta\eta_{\mu\nu})
T^{\mu\nu}_{SM}(\eta_{\mu\nu}) \right]
+{\cal O}(\pi^3).
\label{quadratic}
\end{eqnarray}
where $T^{\mu\nu}_{SM}(\eta_{\mu\nu})$ is the conserved
energy-momentum tensor of the Standard Model evaluated in the
background metric:

\begin{eqnarray}
T^{\mu\nu}_{SM}=-\left.\left(\tilde g^{\mu\nu}{\cal L}_{SM}
+2\frac{\delta
{\cal L}_{SM}}{\delta \tilde g_{\mu\nu}}\right)\right
\vert_{\tilde g_{\mu\nu}=\eta_{\mu\nu}}
\end{eqnarray}

It is interesting to note that there is no single branon
interactions which, as commented above would be related to
Lorentz invariance breaking. In addition the quadratic
expression in (\ref{quadratic}) is valid for any internal $B$
space, regardless of the particular form of the metric
 $\tilde g'_{mn}$. In fact, the form of the couplings only depends  
on the number of branon fields, their mass and the brane tension. 
The dependence on the
geometry of the extra dimensions will appear only at higher orders.

Let us give the results for the different kinds of fields.
The corresponding Feynman rules can be found in Appendix A (we will
follow the notation in \cite{DoMa}):

\subsection*{Scalars}

For a  complex scalar field $\Phi$ with mass $m_{\Phi}$
in a certain representation of a gauge group with generators
$T^a$, the Lagrangian is given by:
\begin{eqnarray}
{\cal{L}}_{S}=(D_{\mu}\Phi)^\dagger
D^{\mu}\Phi-m_\Phi^2\Phi^\dagger\Phi,
\label{scalar}
\end{eqnarray}
whose energy-momentum tensor is given by:
\begin{eqnarray}
T^{\mu\nu}_{S}=-\eta^{\mu\nu}((D_{\rho}\Phi)^\dagger
D^{\rho}\Phi-m_\Phi^2\Phi^\dagger\Phi)+(D^{\mu}\Phi)^\dagger
D^{\nu}\Phi+(D^{\nu}\Phi)^\dagger D^{\mu}\Phi
\end{eqnarray}
where the gauge covariant derivative has the usual form
$D_\mu=\partial_\mu-hA^a_\mu T^a$. From this expression we find
the Feynman rules for the following types of vertices:
$\pi\pi\Phi^\dagger\Phi$, $\pi\pi\Phi^\dagger\Phi A$ and
$\pi\pi\Phi^\dagger\Phi AA$.

\subsection*{Fermions}
The Lagrangian for a Dirac fermion field with
mass $m_\psi$ reads:
\begin{eqnarray}
{\cal{L}}_{f}=\bar\psi (i\gamma^\mu D_\mu-m_\psi)\psi,
\end{eqnarray}
and its associated energy-momentum tensor is:
\begin{eqnarray}
T^{\mu\nu}_{F}&=&\frac{i}{4}\{\bar\psi (\gamma^\mu
D^\nu+\gamma^\nu D^\mu)\psi\
-(D^\nu\bar\psi \gamma^\mu +D^\mu\bar\psi
\gamma^\nu)\psi\}\nonumber\\
&-&\frac{1}{2}\eta^{\mu\nu}\{i(\bar\psi\gamma^\rho D_{\rho}-
D_{\rho}\bar\psi\gamma^\rho)\psi-2m_\psi\bar\psi\psi\}.
\end{eqnarray}
where again we assume the fermion field $\psi$ to be in a
certain representation of a gauge group with generators 
$T^a$, and  the
covariant derivatives are defined as
$D_\mu=\partial_\mu-hT^aA^a_\mu (c_V(a)-c_A(a)\gamma_5)$,
where, in general, the vector and axial couplings could be
different. The
Feynman rules will contain the following types of vertices:
$\pi\pi\bar\psi\psi$ and $\pi\pi\bar\psi\psi A$.

\subsection*{Gauge bosons}

For the Yang-Mills action on the brane we can follow similar
steps. The Lagrangian 
is given by:
\begin{eqnarray}
{\cal{L}}_{A}=-\frac{1}{4}F^{a\,\mu\nu}
F^a_{\mu\nu}+{\cal{L}}_{FP}
\end{eqnarray}
where:
\begin{eqnarray}
F^a_{\mu\nu}=\partial_\mu A^a_\nu-\partial_\nu A^a_\mu-hC^{abc}
A^b_\mu A^c_\nu,
\end{eqnarray}
${\cal{L}}_{FP}$ is the Fadeev-Popov Lagrangian,
which includes the gauge fixing and the ghost terms.
The gauge symmetry can be spontaneously broken in such a way
that the gauge fields acquire  masses $M_a$ through their
couplings to the scalar sector (\ref{scalar}). In that case,
a renormalizable gauge will be used in the calculations.
It is interesting to notice that the metric used in
the gauge fixing term could be
either the induced metric or  the flat background
metric. Both choices give rise to valid  gauge-fixing
expressions. The second
one is however simpler because it does not introduce
new couplings between gauge or Goldstone bosons 
fields and branons. The same criterion has to be
taken for the ghost Lagrangian, which will not be studied here
since we are
only interested in the tree-level analysis.

The energy-momentum tensor takes the form:
\begin{eqnarray}
T^{\mu\nu}_{A}&=&F^a_{\rho\sigma} F^a_{\lambda\theta}(
\eta^{\sigma\lambda}\eta^{\rho\mu}\eta^{\theta\nu} +
\frac{1}{4}\eta^{\rho\lambda}\eta^{\sigma\theta}\eta^{\mu\nu})
+T^{\mu\nu}_{FP}
\end{eqnarray}
and from this interaction action, one finds the Feynman rules for
vertices like $\pi\pi AA$, $\pi\pi A A A$ and $\pi\pi
A A A A$

We see that in all these  vertices,
branons always interact
by pairs with the SM matter fields. In addition, due to
their geometric origin, those interactions are very similar to the
gravitational ones since the $\pi$ fields couple to all
the matter fields through the energy-momentum tensor and with the
same strength, which is suppressed by a $f^4$ factor. In fact, 
branons couple as gravitons do, with the following identification:
\begin{eqnarray}
h_{\mu\nu}&\longrightarrow&-\frac{1}{kf^4}h_{\alpha\beta}(\pi)
\partial_{\mu}\pi^\alpha\partial_{\nu}\pi^\beta,
\end{eqnarray}
where $h_{\mu\nu}$ is the graviton field in linearized gravity.



\section{Constraining brane-world models}

In the brane-world scenario, with $f\ll M_F$, the
relevant low-energy excitations on the brane correspond to the
branons rather than to the Kaluza-Klein graviton excitations. In
such case, and provided the energy scale of present accelerators
is well below both the fundamental scale of gravity $M_F$ and the
brane tension scale $f$, but is large compared to the branon mass
$M$, such branon fields will be produced in the collision of SM
particles. Moreover,  the calculation of the corresponding
production cross-sections could be performed by using the
effective theory described in the previous section.

In principle, there would be new physics related to both
the KK gravitons  and the branons, but the
dependence on $f$ in each case is completely different. For example, the
behavior of typical cross sections for missing energy
processes, i.e. KK graviton and branon production processes,
becomes large for both small and large values of the brane tension
$f$ \cite{Kugo}.
The cross section for a small value of $f$ is governed by the
branon production process, while for a large value, the KK
gravitons production rate dominates.
Therefore if the brane tension scale is smaller than
the fundamental gravitational one, then the first
indications of extra dimensions would be given by the production of
branons. This would allow us to measure the brane tension and the branon mass
or, if branons are not found, at least to set bounds on such parameters.

The  effective couplings introduced in the previous section
provide the necessary tools to compute the cross sections and the
expected rates of new exotic events in terms of $f$ and the
branons masses only. In fact, with a rotation in the coset space
$K$, the branon mass matrix takes a diagonal form. We will assume
that there are $n$ branons with the  same mass $M$. If the
branons masses are similar this would be a good approximation.
In the opposite case, with very different masses,
we could neglect the heavier fields and consider only the
production of light branons. In these simple cases,
all the cross sections will be parametrized by
three unknown parameters:


\begin{center}
$n$: number of branons.

$M$: mass of the branons.

$f$: brane tension scale.
\end{center}

In the following, we will compare the data coming from LEP
with the predicted cross section
for different processes in which we expect
high sensitivity to new physics. In particular we will concentrate
on the branon contribution to the invisible width of the
Z boson, to the decay rate of the $W^{\pm}$
bosons and to direct searches with single-photon and single-Z 
vertices. This  will allow us to set bounds on certain
combinations
of the above parameters.
\section{Bounds from the Z invisible width}

In the Standard Model, the full decay width of the Z
boson has three types of contributions, coming from charged leptons,
hadrons
and neutrinos respectively. The third one corresponds to the
so called invisible width, since neutrinos escape without producing
any signal in the detectors. In principle, physics
beyond the SM could also give rise to invisible decays
of Z and therefore
to deviations with respect to the SM predictions.

The precision measurements done mostly at LEP-I set stringent
limits on the Z invisible decay width, which
can be translated into strong bounds on the new channels contributing
to such process. In fact in the following, we will show how it is
possible to get  lower bounds
on the brane tension $f$ coming from light branons ($M<M_Z/2$) in
which
the Z could  decay. Since branons are stable and chargeless
particles, they would also avoid detection.

In this section we calculate the first order contribution from
branons to this width, which is given by the decay of Z into two
branons and two neutrinos:

\begin{center}
$\Gamma_Z^b$:
$Z\longrightarrow\bar\nu(p_1),\nu(p_2)\pi(k_1),\pi(k_2)$
\end{center}

We use the expression for the probability amplitudes in Appendix B
Eq.(\ref{Z0amp}) calculated with the Feynman rules given in
Appendix A. We integrate over the branon two-body phase
space making use of the formulae in Appendix C. The result is
averaged over the three Z polarizations. In addition, such
expression should be divided by two because the outgoing branons
are indistinguishable. Thus we get:


%

\begin{eqnarray}
\frac{d\Gamma_Z^b}{d^3\overrightarrow{p_1}d^3\overrightarrow{p_2}}&=&
\frac{\vert h\vert^2}{4\pi} \frac{2M_Z n
}{61440f^83\pi^6t^2u^2(M_Z^2-t)(M_Z^2-u)(M_Z^2-s)^2}
\nonumber\\
&&\sqrt{1-\frac{4M^2}{k^2}}
\{20M^2 M_Z^2(2k^2-5M^2)t^2u^2(M_Z^2(2s-k^2)+tu)\nonumber\\
&+&(k^2-4M^2)^2\{stu(s(k^2+M_Z^2)+4tu)(2s(k^2+M_Z^2)+t^2+u^2)
\nonumber\\
&+&(t^2+u^2)(2s+2k^2+M_Z^2)M_Z^8\nonumber\\
&-&[6s(t^3+u^3)+6s^2(t^2+u^2-tu)+3tu(t-u)^2+t^4+u^4]M_Z^6\nonumber\\
&+&M_Z^4[2s^3(2(t^2+u^2)-5tu)+2s^2(3(t^3+u^3)-5tu(t+u))\nonumber\\
&+&s(2(t^4+u^4)+tu(t^2+u^2-8tu))+tu(t^3+u^3-7tu(t+u))]\nonumber\\
&-&M_Z^2[s^4(t-u)^2-8t^3u^3+2s^3(t^3+u^3-2tu(t+u))\nonumber\\
&+&2stu(t^3+u^3-3tu(t+u))\nonumber \\
&+&s^2(t^4+u^4+tu(t^2+u^2-14tu))]
\}\}
\label{Z0decay}
\end{eqnarray}
where we have used equations (\ref{Z0amp})
with $c_V=c_A=1$, $M_a=M_Z$
and the coupling constant $h=g/(4\cos \theta_W)$, with
$\theta_W$ the Weinberg angle. We have  also
  made use of the Mandelstam variables:
$s\equiv(p_1+p_2)^2$, $t\equiv(p_1-q)^2$, $u\equiv(p_2-q)^2$ and
$k^2\equiv(k_1+k_2)^2$, although only three of them are independent
because $s+t+u=k^2+M_Z^2$.

We can also compute the above differential cross
section in terms of (non-covariant) variables with a more physical
interpretation, namely, the energies of the outgoing
neutrino and  antineutrino, 
$p_1^0\equiv
E_1$ and $p_2^0\equiv E_2$ respectively, and the angle between
their three-momenta $\theta$. In terms of these variables we have:
\begin{eqnarray}
s&=&2p_1^0p_2^0(1-\cos \theta),\nonumber\\
t&=&M_Z(M_Z-2p_1^0),\nonumber\\
u&=&M_Z(M_Z-2p_2^0),
\label{mandelstam}
\end{eqnarray}
 The total width is calculated by  integrating  the above
expression with respect to the three variables within  the
following kinematic limits:
\begin{eqnarray}
p_1^0&\in&I_1\equiv[0,\frac{M_Z^2-4M^2}{2M_Z}],\\
p_2^0&\in&I_2\equiv[0,\frac{M_Z(M_Z-2p_1^0)-4M^2}{2(M_Z-(1
-\cos\theta)p_1^0)}],\nonumber\\
\cos \theta&\in&I_0\equiv[-1,1].\nonumber
\label{limits}
\end{eqnarray}
The final result, depending on the three undetermined parameters
reads:
\begin{eqnarray}
\Gamma_Z^b(n,f,M)=\int_{I_0\times I_1\times
I_2}\frac{d\Gamma_Z^b}{dp_1^0 dp_2^0d\cos \theta}dp_1^0
dp_2^0d\cos \theta.
\label{final}
\end{eqnarray}
The integration over the rest of  angles has been done immediately
because the decay rate  does not depend of them.
It is interesting to note that the dependence on the number
of branons $n$
and brane tension $f$ factorizes, and only
the dependence on $M$ is non-trivial:
\begin{eqnarray}
\Gamma_Z^b(n,f,M)=\frac{n M_Z^9}{f^8}\times \Pi(M/M_Z)
\label{totalZ0}
\end{eqnarray}
where $\Pi(x)$ is a dimensionless function, which depends on a single
dimensionless variable, and which can be easily calculated from
(\ref{final}). For example, for massless
branons a numerical integration gives:
\begin{eqnarray}
\Gamma_Z^b(n,f,0)=\frac{n M_Z^9}{f^8}\times 6.15\times10^{-13}.
\end{eqnarray}

\begin{figure}[h]
{\epsfxsize=12.0 cm \epsfbox{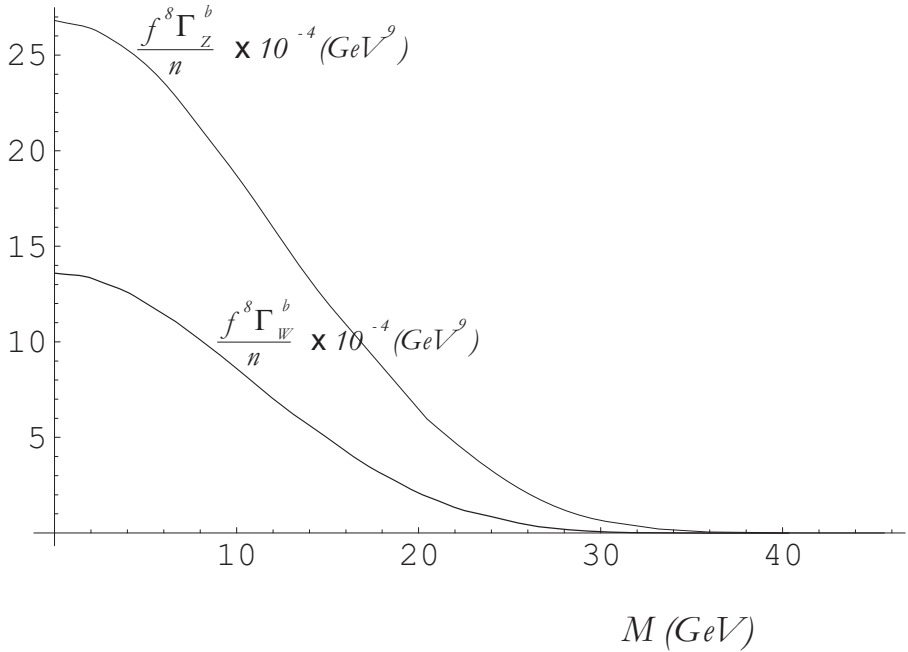}}
\caption{\footnotesize{$W^\pm$ and Z
widths as a function of the branon mass. Both plots correspond
to a single channel. We have extracted the 
dependence
on the brane tension and the number of branons in the factor $f^8/n$.}}
\end{figure}

To show the dependence on the branon mass, we  plot in the Figure
1, $\Gamma_Z^b(n,f,M)f^8/n$ against $M$.
Finally, to get the total Z invisible width into branons,
we have to multiply by a factor of 3 because there are three different
neutrino families. 

The key observation is that at the 95$\%$
confidence level, the
variation of the Z invisible width cannot be larger than 2.0 MeV
\cite{LEP}, i.e., the contribution from new physics satifies:
\begin{eqnarray}
\Gamma_Z^{new}(n,f,M)\,<\, 2.0 \,\mbox{MeV},
\end{eqnarray}
It is then possible  to find bounds for the different
brane parameters just imposing the above limit
on Eq.(\ref{totalZ0}). For example, if branons are massless,
the bound on $f$ depends only on the number of branons:
\begin{eqnarray}
f\,>\, 11.9\, n^{1/8}\, \mbox{GeV}.
\end{eqnarray}

On the other hand, in the case of  one  branon, we show the
exclusion plot in the $f-M$ plane (see Figure 2).
\begin{figure}[h]
{\epsfxsize=12.0 cm \epsfbox{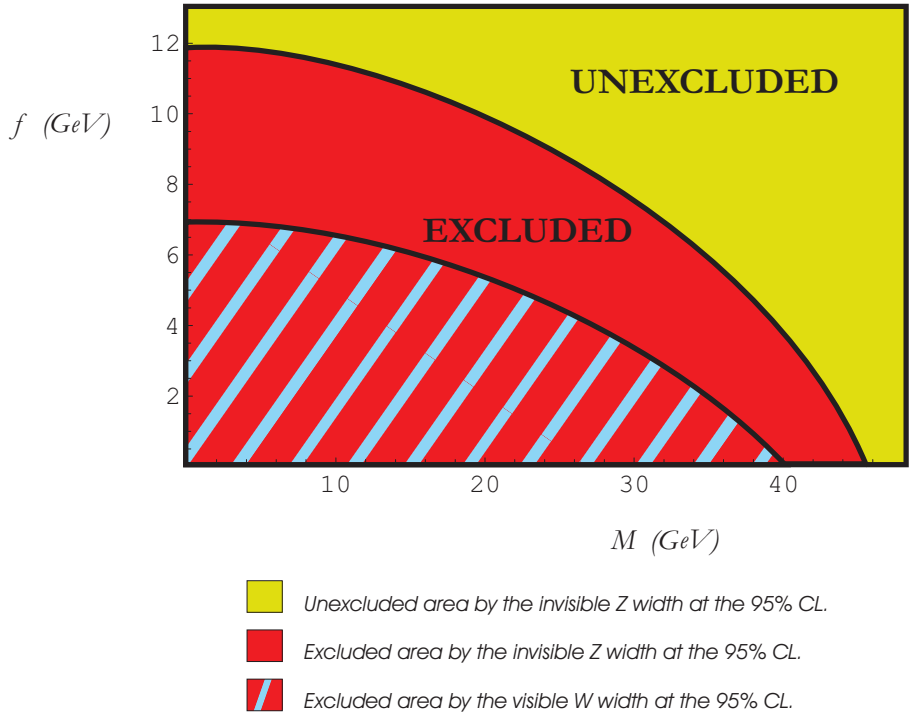}}
\caption{\footnotesize{Exclusion plot in the $f-M$ plane
for single-branon models from LEP-I data.
The dark area is excluded by the measurement of the Z
invisible width and the striped region is excluded by the measurement
of the visible $W^\pm$ decay width.}}
\end{figure}

\section{Bounds from $W^\pm$ decay}
There are mainly two channels contributing to this process.
First, the decay of $W^\pm$ into two branons
and two leptons:
\begin{center}
$\Gamma_W^{b}$: $W^-\longrightarrow
l^-(p_1),\bar\nu(p_2),\pi(k_1),\pi(k_2)$
\end{center}
Such leptons can be an electron and an electron antineutrino for
the $W^-$ decay, their antiparticles for $W^+$, or the analogue
pairs of leptons in the rest of families. The results of the
different decay rates agree in the limit in which the lepton
masses are negligible. In fact this is a good approximation
 since the typical energy carried by the of outgoing
particles will be comparable to the $W^\pm$ mass which is
much larger than the leptons masses.

Second, we have the $W^\pm$ decay into 
two branons and a quark-antiquark pair:
\begin{center}
$W\longrightarrow
q(p_1),\bar q(p_2),\pi(k_1),\pi(k_2)$
\end{center}
In principle there should be also a channel including gluons 
in the final state, however in the SM, such channel is 
suppressed by a coefficient  $\alpha_S(M_W)/\pi=0.04$ and 
for that reason we will also ignore it
in the present case. In any case, the result we will obtain
will be a strict lower bound to the true decay width into branons. 

The calculation is totally similar to that of the Z decay
and for that reason we will not repeat it here.
 The expressions for the Z decay in
(\ref{Z0decay}) are also
valid for the $W^\pm$, the only changes, apart from the
the different gauge-boson mass,  are the $W^\pm$ couplings
to leptons and quarks. Thus, in this case we have:
$M_a=M_W$, $c_V=
c_A =1$, 
$h=g/(2\sqrt{2})$ for leptons and 
$h=V_{ij}g/(2\sqrt{2})$ for quarks where  $V_{ij}$
is the Cabbibo-Kobayashi-Maskawa (CKM) matrix element with
$i=u,c,t$ and $j=d,s,b$.
The differential decay rate has the same form as in the
Z case, which is written again in terms of the energy of the
two leptons or the quark-antiquark pair $p_1^0$, $p_2^0$ 
and the angle between their three momenta
that we denote by $\theta$. Thus the corresponding Mandelstam
variables are obtained from (\ref{mandelstam}) just
replacing $M_Z\rightarrow M_W$
The total integration over the leptons or quarks phase-space
has  to be done in a numerical form, with the analogous
limits to those in (\ref{limits}) again replacing the
gauge bosons masses.

The final result $\Gamma_W^{b}(n,f,M)$
depends on the three unknown parameters and is obtained from
the differential rate by using the analogue expression
to (\ref{final}).
We see that once again, the dependence on $n$ and $f$ factorizes,
and  the
dependence on $M$ enters through the same dimensionless
function $\Pi(x)$. Thus we can write for the leptonic decay:
\begin{eqnarray}
\Gamma_W^{b}(n,f,M)=2\cos^2\theta_W
\frac{n M_W^9}{f^8}\times \Pi(M/M_W).
\label{leptonic}
\end{eqnarray}
For massless branons we obtain by  numerical integration:
\begin{eqnarray}
\Gamma_W^{b}(n,f,0)=\frac{n M_W^9}{f^8}\times 9.56\times10^{-13}.
\end{eqnarray}
The results in the case of quarks can be obtained 
in a straightforward way from the
leptonic one just multiplying by the  modulus
squared of the corresponding CKM matrix element.

To show the dependence on the branon mass, we have plotted 
$\Gamma_W^{b}f^8/n$ 
against the branon mass $M$
in Figure 1.

The most important difference with respect to the Z analysis
arises from  the experimental constraint on the $W^\pm$ width
coming from new physics. In this case, the decay is visible and,
for example, if the final decay products contain $e^-$ and
$\bar\nu_e$, the process with branons could give a signal similar
to the following Standard Model process: $\Gamma_W^{SM}$:
$W^-\longrightarrow
e^-(p_1),\bar\nu_e(p_2),\nu_\tau(p_3),\bar\nu_\tau(p_4)$, in
which, the $W^-$ decays into a $\tau^-$ and a $\bar\nu_\tau$, and
then the $\tau^-$ decays finally into an $e^-$, $\bar\nu_e$ and
$\nu_\tau$.

The behaviour of the $W^\pm$ decay width in
terms of the visible lepton energy  $E_l$ is plotted in Figure 3
for different branon masses.
\begin{figure}[h]
{\epsfxsize=12.0 cm \epsfbox{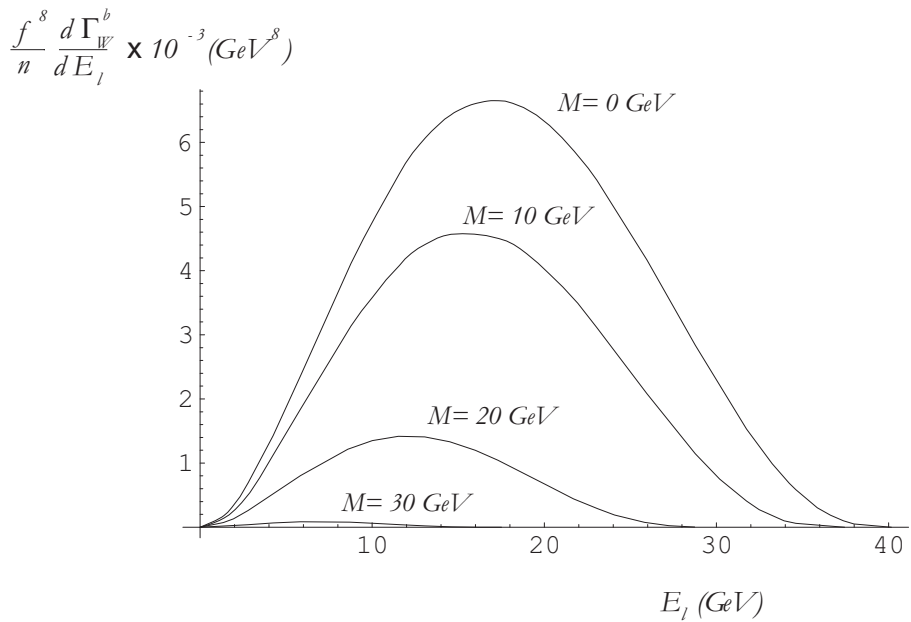}}
\caption{\footnotesize{Behaviour of the $W^\pm$ decay width in
terms of the visible lepton energy  $E_l$ for different branon
masses.}}
\end{figure}

However in this work, we are only interested in the constraints on
the theory parameters coming from the total $W^\pm$ width. 
The total leptonic decay is obtained multiplying (\ref{leptonic}) 
by a factor of 3, because 
there
are three different kinds of processes corresponding to the three
leptonic families. On the other hand, to calculate the hadronic decay 
we have  to multiply by:
\begin{eqnarray}
3\sum_{i=u,\,c\,;j=d,\,s,\,b}|V_{ij}|^2= 6.117\pm 0.075,
\end{eqnarray}
where the factor of 3 comes from the numbers of colors and the sum 
is extended to 
the six CKM matrix elements which do not 
involve the top quark. The numerical result 
is taken from the combined LEP measurement of 
$Br(W\longrightarrow q\bar q)$ \cite{W}.
The above approximation assumes that the mass of these five quarks is 
neglected when compared to the  typical energy of the process ($M_W$), 
and the top contribution is neglected because of its large mass.
We can thus estimate the total $W^\pm$ width just 
multiplying  by $3+6.12$ 
the result in (\ref{leptonic}). 

 In order to get bounds on the contribution from new physics to the
$W^\pm$ width, $\Gamma_W^{new}$, we take into account
the uncertainties 
on the experimental measurement of the visible W width: 
$\Gamma_W^{LEP}=2.150 \pm 0.091$ GeV \cite{LEP}, and the 
uncertainties in the SM calculation:
$\Gamma_W^{SM}=2.093 \pm 0.003$ GeV \cite{W}. So 
that for the total
allowed variation in the visible $W^\pm$ width
we obtain at the $95\%$ confidence level:
\begin{eqnarray}
\Gamma_W^{new}(n,f,M)\,<\, 
(\Gamma_W^{LEP}-\Gamma_W^{SM})+2\Delta\Gamma_W^{LEP}=240 \,\mbox{MeV},
\end{eqnarray}
This limit translates into the following one for massless branons:
\begin{eqnarray}
f\,>\, 6.9\, n^{1/8}\, \mbox{GeV}.
\end{eqnarray}

The exclusion regions as a function of the variables $f$ and $M$ are shown 
in Figure 2 for the single branon case.

\section{Branon bounds from direct searches}

\subsection{Cross section with a generic gauge boson}

In this section we present the cross section of the process:
\begin{eqnarray}
\sigma_A:
\bar\psi(p_1),\psi(p_2)\longrightarrow\pi(k_1),\pi(k_2),A^a_\mu(q)
\end{eqnarray}
The initial particles can be
either leptons or quarks, although we will be
mainly interested in the $e^+e^-$ case. There
are three final particles, one arbitrary gauge boson (either
Z or $\gamma$)
together with the two branons. We are interested in the
differential cross section with respect to the gauge boson phase-space
parameters, i.e. energy  and scattering angle.

We have integrated over the two-body phase space of branons. The
probabilities are calculated as usual by doing the
square of the amplitude module given in (\ref{Z0eamp}). We have also
averaged over the spin of the
initial particles and summed over polarizations
 of the outgoing
boson. The
probability has to be divided by two because the outgoing
branons are indistinguishable.

We define the invariants: $s\equiv(p_1+p_2)^2$,
$t\equiv(p_1-q)^2$,$u\equiv(p_2-q)^2$ and $k^2\equiv(k_1+k_2)^2$.
Again only three of them are independent because $s+t+u=k^2+M_a^2$
($M_a$ is the gauge boson mass and we are neglecting the fermion
masses). In the CM system the probability should be divided by
$2p_1^0 2p_2^0 |v_1-v_2|=2s$.

Again we can change to more physical variables,
namely, the energy fraction of the emitted gauge
boson $x=2q_0/\sqrt{s}$
and the scattering angle $\theta$. Thus we get:
\begin{eqnarray}
t&=&-\sqrt{s}(q_0-|\overrightarrow{q}|\cos(\theta))+M_a^2\nonumber\\
u&=&-\sqrt{s}(q_0+|\overrightarrow{q}|\cos(\theta))+M_a^2\nonumber\\
k^2&=&s(1-x)+M_a^2\\
q_0&=&\frac{x\sqrt{s}}{2}\nonumber\\
|\overrightarrow{q}|&=&\frac{1}{2}\sqrt{sx^2-4M_a^2}\nonumber
\end{eqnarray}
and the phase-space volume takes the form:
\begin{eqnarray}
\frac{d^3q}{2q_0(2\pi)^3}&=&\frac{dk^2dtd\phi}{4s(2\pi)^3}
=\frac{dxd(\cos\theta)d\phi}{8(2\pi)^3}\sqrt{(sx)^2-4sM_a^2}.
\end{eqnarray}
Therefore, for $M_a\neq 0$ and arbitrary $c_V$ and $c_A$
parameters, we have:
\begin{eqnarray}
\frac{d\sigma_A}{dxd\cos\theta}&=&
\frac{\vert h\vert^2}{4\pi}
\frac{(c_V^2+c_A^2)n}{122880f^8\pi^2(s-M_a^2)^2}
\sqrt{1-\frac{4M^2}{s(1-x)+M_a^2}}\nonumber\\
%
&&\sqrt{s(sx^2-4M_a^2)}\{5M^2 M_a^2(2(s(1-x)+M_a^2)-5M^2)\nonumber\\
&&\{8M_a^2+(sx^2-4M_a^2)\sin^2\theta\}\nonumber\\
%
%
&+&\frac{2(s(1-x)+M_a^2-4M^2)^2}{(4M_a^2(M_a^2+s(1-x))
+s(sx^2-4M_a^2)\sin^2\theta)^2}
\nonumber\\
&&\{16M_a^2(s(1-x)+M_a^2)^2[10M_a^6+M_a^4s(9-5x)\nonumber \\
&+&2M_a^2s^2(4-3x+x^2)+s^3(1-x)]\nonumber\\
&-&2[40M_a^{12}-2M_a^{10}s(-112+40x+5x^2)\nonumber \\
&+&4M_a^8s^2(78-70x-3x^2+5x^3)+s^6x^2(-2+4x-3x^2+x^3)\nonumber\\
&+&M_a^6s^3(216-272x+6x^2+66x^3-11x^4)\nonumber\\
&+&M_a^4s^4(96-184x+62x^2+44x^3-21x^4+x^5)\nonumber\\
&+&M_a^2s^5(8-16x-12x^2+42x^3-29x^4+6x^5)]\sin^2\theta\nonumber\\
&+&s(sx^2-4M_a^2)^2[7M_a^6+M_a^4s(17-7x)+M_a^2s^2(1-2x)\nonumber\\
&+&s^3(3(1-x)+2x^2)]\sin^4\theta-s^2(sx^2-4M_a^2)^3
(s-M_a^2)\sin^6\theta\}\}.\nonumber
\label{total}
\end{eqnarray}
Notice that this expression is also applicable to quark-antiquark
collisions just by choosing the appropriate couplings ($c_V$ 
and $c_A$ parameters together with $h$).




Also, for certain processes it is interesting to calculate the
cross sections without summing over the polarizations ($\epsilon_\mu^a$)
of the outgoing gauge boson. Again we consider the case $M_a\neq0$
and arbitrary
$c_V$ and $c_A$ parameters. The results can
be found
in Appendix D for longitudinal (\ref{long})
and transverse (\ref{trans}) polarizations. 
It is interesting to note  that the longitudinal
cross secction vanishes in the limit in which the gauge boson mass 
goes to  zero.
Therefore, the total cross section (including the
three polarizations) in the limit
$M_a\longrightarrow 0$ has the same form as the total cross
section for a massless gauge field ($M_a=0$), and it reads:
\begin{eqnarray}
\frac{d\sigma_A}{dxd\cos\theta}&=&\frac{\vert h\vert^2}{4
\pi}\frac{s(c_V^2+c_A^2)
(s(1-x)-4M^2)^2 n}{61440f^8\pi^2}\sqrt{1-\frac{4M^2}{s(1-x)}}
\nonumber\\
&&\left[x(3-3x+2x^2)-x^3\sin^2\theta
+\frac{2(1-x)(1+(1-x)^2)}{x\sin^2\theta}\right]\nonumber \\
\end{eqnarray}
As expected, the cross section is divergent for
collinear outgoing boson ($\theta=0$)  and
also for a soft ($x=0$) massless gauge
boson.

\subsection{Bounds from single-photon processes}

The differential cross section obtained
in the previous section could be  used in
direct searches of branons in colliders. In this section 
 we are going to calculate the contribution to the
total cross section of processes involving a single
photon in the final state. This will allow us to obtain new
bounds on $M$ and $f$ assuming non observation at LEP-II.

To perform the angular integration  over the polar angle 
$\cos\theta$ in (\ref{total}),
it is necessary to take into account the
angular range covered by the
detector $\cos\theta \in
[-d,d]$. Thus we get:
\begin{eqnarray}
\frac{d\sigma_\gamma}{dx}&=&\frac{\vert h\vert^2}{4
\pi}\frac{s(c_V^2+c_A^2)(s(1-x)-4M^2)^2 n}{184320\,
    f^8\,{\pi }^2\,x}\sqrt{1-\frac{4M^2}{s(1-x)}}\\
&&\left( 2\,d\,x^2\,\left( 9
-9\,x + \left( 3 + d^2 \right) \,x^2 \right)\right.\nonumber \\
&+&\left.3\,\left( -2 + 4\,x - 3\,x^2 + x^3 \right) \,
       \log \left(\frac{{\left( 1 - d \right) }^2}{{\left( 1
 + d \right) }^2}\right)
       \right)\nonumber.
\end{eqnarray}
where $h=e$, $c_V=1$ and $c_A=0$.
%
%
\begin{figure}[h]
{\epsfxsize=12.0 cm \epsfbox{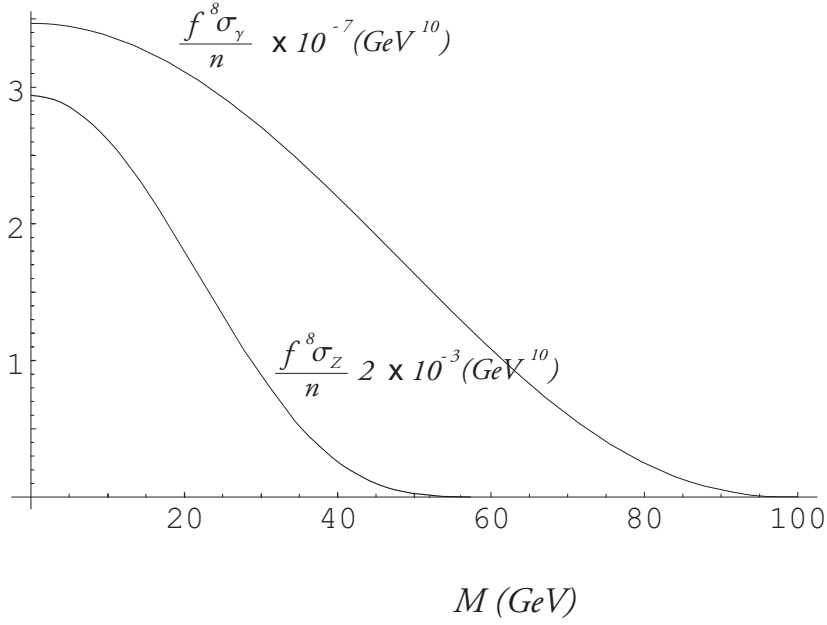}}
\caption{\footnotesize{Total cross sections for single-photon and
single-Z processes as a function of the branon mass. The dependence on the
brane tension and the number of branons has been factorized out
in the $f^8/n$ coefficient.}}
\end{figure}

The integration in the $x$ variable is done within the following 
limits $x
\in [E_m/(2\sqrt{s}),1-4M^2/s]$. The upper bound is
fixed by kinematical constraints, but the lower one ($X=E_m/2\sqrt{s}$), 
imposing a minimum energy for the photons ($E_m$), depends on specific
experimental conditions (trigger, noise, backgrounds, \ldots).
To perform the numerical study we take $d=0.96$ and $E_m= 5$ GeV, which 
correspond to the typical boundary conditions of one of the 
LEP experiments (see  \cite{L3}). We will also use $\sqrt{s}=
206$ GeV to do the calculations. Although the maximum centre-of-mass
energy was close to 209 GeV, most of the data  at LEP-II 
taken in the last year of running
were collected in the range between 205 and 207 GeV \cite{LEP}.

The total cross section can be calculated analytically, although
it will not be shown here since the final
expression is quite long. The result for massless branons (M=0) is
much simpler, being just proportional to $s^3$:
\begin{eqnarray}
\sigma_\gamma(M=0)&=&\frac{n A_\gamma s^3}{f^8}.
\label{zeromass}
\end{eqnarray}
where the  constant depends on the detection limits as follows:
\begin{eqnarray}
A_\gamma&=&\frac{-\alpha_{EM}}{11059200\,{\pi }^2}\left[ \,\left[ 2\,d\,{
\left( -1 + X \right) }^3\,
         [ 30\,\left( 1 + 3\,X - 3\,X^2 + X^3 \right)\right.\right.
         \nonumber\\
         &+&\left.
           d^2\,\left( 1 + 3\,X + 6\,X^2 + 10\,X^3 \right)  \right]
          +
        3\,\log \left(\frac{{\left( d -1 \right) }^2}{{\left( 1
+ d \right) }^2}\right)\left( -247
+ 480\,X\right.
        \nonumber\\
         &-&\,
        \left.\left.\left.   390\,X^2 + 220\,X^3 - 75\,X^4
+ 12\,X^5 - 120\,\log (X) \right)
        \right]\right]
         %
\end{eqnarray}
with $\alpha_{EM}=e^2/(4 \pi)$.
\begin{figure}[h]
{\epsfxsize=12.0 cm \epsfbox{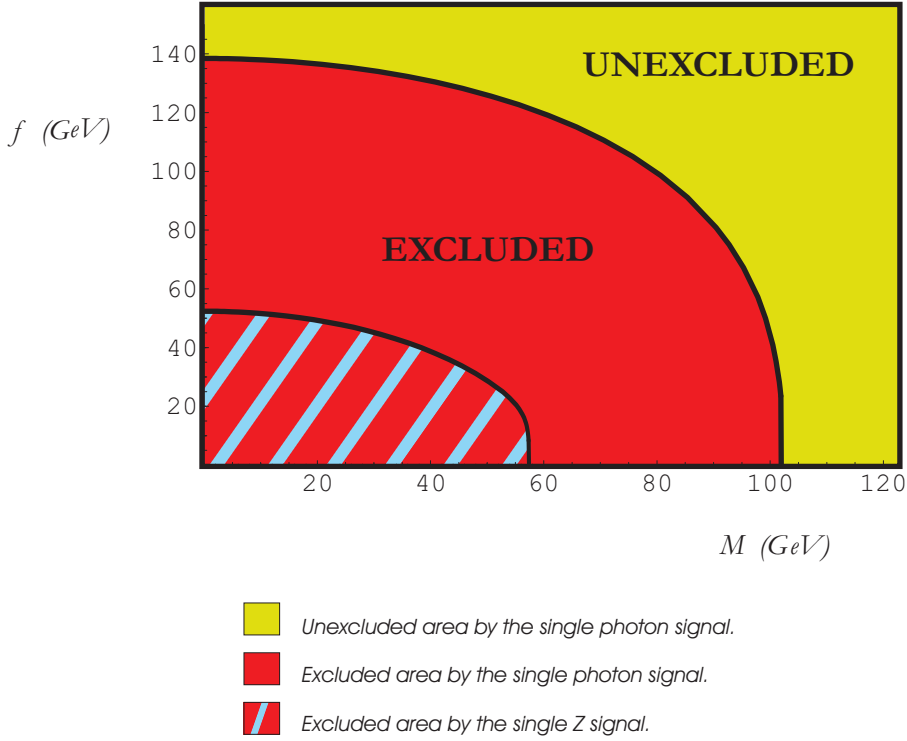}}
\caption{\footnotesize{Exclusion regions in the $f-M$ plane
coming from single-photon events (dark region) and single-Z
processes (striped area), using LEP-II data.}}
\end{figure}

For the limits mentioned before we get $A_\gamma=4.54\times 10^{-7}$. The
branon mass dependence of the total cross section is plotted in
Figure 4.

To obtain the bound on the theory parameters, we can consider the
limit
on the total cross section coming from new physics
which could be added to the single
photon channel without being detected. For LEP-II, this limit is
approximately $\sigma_{new}\simeq 0.1$ pb, which is the typical
experimental sensitivity observed in LEP searches \cite{higgs_searches}. 
The result
for one branon assuming no observation is plotted in Figure 5.
Thus the interior area limited by the bound curve is potentially excluded 
by the LEP-II experiment. In the case of massless branons, the limit
depending on the number of branons reads:
\begin{eqnarray}
f \geq 138 n^{1/8} \mbox{GeV}.
\end{eqnarray}
\begin{figure}[h]
{\epsfxsize=12.0 cm \epsfbox{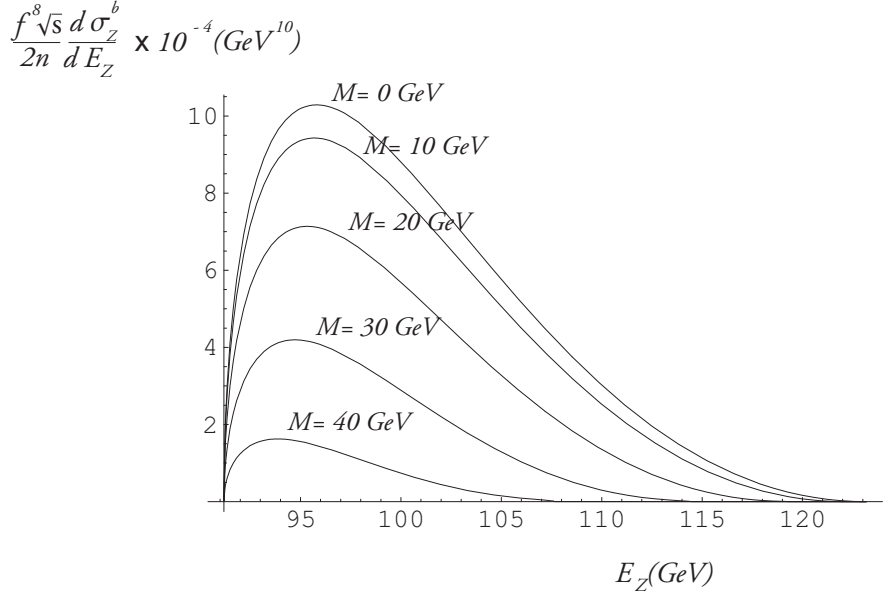}}
\caption{\footnotesize{Differential cross section for single-Z
processes as a function of the outgoing Z energy for
different branon masses and for a center of mass energy $\sqrt s=206$
GeV. The dependence on $f$ and $n$ has been factorized out. }}
\end{figure}

\subsection{Bounds from single-Z processes}

As in the single-photon case, the single-Z channel can
be used to restrict the
parameters of the theory. In this section we are going to
calculate the total cross section summing over Z
polarizations to estimate these bounds.

Again to implement the angular integration over the polar angle
$\cos\theta$, we take into account the domain of the
detector: $\cos\theta \in [-d,d]$, with $d=0.96$. The behavior of
the cross section in terms of the outgoing Z energy is
represented in Figure 6 for different branon masses (for
$\sqrt{s}=206$ GeV). In this case, $h=g/(4 \cos \theta_W)$, 
$g=e/\sin \theta_W$, $c_V=-1+4\sin^2 \theta_W$ and $c_A=1$.
%
%

The integration in the $x$ variable  is done within the following 
limits $x
\in [M_Z/(2\sqrt{s}),1+(M_Z^2-4M^2)/s]$, where both
the lower and the upper limits are kinematical.

To estimate a bound on the theory parameters, we will consider
again the
 limit on the total cross section coming from new
physics that can be added to the
single Z channel without being detected.
For LEP-II, the order of magnitude of this limit is
approximately equal to that in the single photon case
$\sigma_{new}\simeq 0.1$ pb.

The result for massless branons is:
\begin{eqnarray}
f \geq 52 n^{1/8} \mbox{GeV}.
\end{eqnarray}
and the exclusion plot can be found in Figure 5.
The interior area limited by the bound curve, which is excluded, is
smaller than in the single photon analysis. 
This is an expected result since the Z
coupling to
the electron field is smaller than that of the photon,
and the Z mass restricts the avalaible phase space.
In fact, this restricts the search only to branons with
masses:
$M\leq(\sqrt{s}-M_Z)/2\simeq 57$ GeV. In the single-photon
channel however, the kinematical range is larger
$M\leq\sqrt{s}/2\simeq 103$ GeV.
Despite this fact, the single Z channel analysis
is still interesting since it provides
a completely independent direct search method.
In  future colliders, working at very high
centre-of-mass energy, the Z mass
could be neglected $\sqrt{s}>>M_Z$, and the total cross section for the
two channels  will take the same form.  The
only difference will come from the ratio of couplings to the
electron field, which is:
\begin{eqnarray}
\frac{\sigma_Z}{\sigma_\gamma}(\sqrt{s}>>M_Z)
\simeq\frac{(4\sin^2{\theta_W}-1)^2+1}{16
\cos^2{\theta_W}\sin^2{\theta_W}}\simeq 0.37,
\end{eqnarray}
where we have use the value $\sin^2{\theta_W}= 0.22$. This implies
that provided the rest of the analysis remains unchanged, the
bounds on $f$ which come from these two processes are similar:
\begin{eqnarray}
\frac{f(\sigma_Z)}{f(\sigma_\gamma)}(\sqrt{s}>>M_Z)\simeq
(0.37)^{1/8}\simeq 0.88,
\end{eqnarray}

Although, in general,  the single-Z channel has more
background compared to the
single photon case, implying lower precision, however it
allows us to perform analysis
depending on the polarization which could improve the bounds.

Since branon effects grow strongly with energy, it
 is not surprising that the bounds from direct searches
with  ($\sqrt{s}\simeq 200$ GeV) are more constraining
than the indirect ones, in which the energy scale
is set by the Z mass
  ($M_Z\simeq 90$ GeV). However, because this is an effective
theory the growth with energy of the cross sections will eventually
violate unitarity. At that point the approximation will no longer be
valid.
It is also interesting to note
that the present bounds improve the astrophysical
ones for massless branons, 
coming from the 1987a supernova, from which $f > 10$ GeV \cite{Kugo}.

\section{Prospects for future linear colliders}

 Several proposals for the construction of  $e^+e^-$ linear 
colliders in the TeV range are currently under study. The TESLA 
(TeV Energy Superconducting Linear Accelerator) \cite {TESLA}, the NLC 
(Next Linear Collider) \cite{NLC} and the JLC (Japanese Linear Collider) 
\cite{JLC} are examples of the first generation of these colliders, whereas
the CLIC (Compact Linear Collider) \cite {CLIC} would correspond to the
second generation.  In this section we discuss the sensitivity of these 
colliders to a hypothetical branon signal. The study will be performed 
in terms of the brane tension, $f$, and the branon mass, $M$.

  The physics programme of the new linear collider projects includes
the measurement of electroweak parameters with improved precision, 
such as the 
invisible Z width or the $W^\pm$ width. 
However, and since the deviations due to the 
presence of branons increase dramatically with energy, the 
largest sensitivity 
to a branon signal is expected in direct searches like 
single-$\gamma$ and 
single-Z. 

  In order to estimate the sensitivity of future linear colliders to branon 
signals, the LEP-II study of the previous section is extended 
to higher center-of-mass energies. It is assumed that, at the time of
construction of these accelerators, the theoretical and systematic 
uncertainties on Standard Model processes will be controlled at the level 
of the femtobarn. Under this assumption, we will estimate the sensitivity limit,
$\sigma^{FLC}_{new}$, by scaling the LEP-II estimate by the expected 
gain in statistics:
\begin{eqnarray}
\sigma^{FLC}_{new}=\sqrt{\frac{(TIL_{LEP-II})}{(TIL_{FLC})}}\,
0.1~\mbox{pb}.
\end{eqnarray}

\noindent
where $TIL_{LEP-II} \simeq 700~\mbox{pb}^{-1}$ is the LEP-II integrated 
luminosity and $0.1\,\mbox{pb}$ is the LEP-II 
sensitivity limit. We will 
consider the following values of the integrated luminosity at future 
colliders, $TIL_{FLC}$: $TIL_1\simeq 200$ \mbox{fb}$^{-1}$ for a first stage 
of TESLA, NLC or JLC, and $TIL_2\simeq 1000$ \mbox{fb}$^{-1}$ as a maximum 
value for a second stage. For CLIC we will assume the same
total integrated luminosity: $TIL_{CLIC}=TIL_2\simeq 1000$ \mbox{fb}$^{-1}$. The 
cross section bounds for both luminosity choices are similar:
$\sigma_{new}^{1}=6$ fb  and $\sigma_{new}^{2}=\sigma^{CLIC}_{new}=3$ fb.

  The critical parameter in the analysis is the center-of-mass energy,
$E_{CM}$. In the single photon channel, this leads to limits
on $f$ for any branon mass $M<E_{CM}/2$. In particular the limit  
for massless branons ($f_0$) increases proportionally to 
$E_{CM}^{3/4}$ according to Eq.(\ref{zeromass}). Assuming
a center-of-mass energy for the first stage of the first generation of
linear collider of approximately $500$ GeV, $1$ TeV for its second stage and 
$5$ TeV for CLIC, we obtain the following 
limits for
$f_0$ in the case of a single branon: $f_{0\,\gamma}^{1}>398$ GeV, 
$f_{0\,\gamma}^{JLC}>758$ GeV and $f_{0\,\gamma}^{CLIC}>2.64$ TeV.           
The results of a full study in the $(f,M)$ plane and in different 
experimental contexts are presented in Figure 7.

For the single-Z channel, the bounds are less restrictive:
$f_{0\,Z}^{1}>205$ GeV, 
$f_{0\,Z}^{2}> 450$ GeV an $f_{0\,Z}^{CLIC}>1.87$ TeV. Obviously, 
the study
in this case is only applicable to branon masses below 
$(E_{CM}-M_Z)/2$, due 
to kinematic constraints. The excluded regions in 
the $(f,M)$ plane are also shown in Figure 7.

\hspace{1.5 cm}
{\epsfysize=18.3 cm \epsfbox{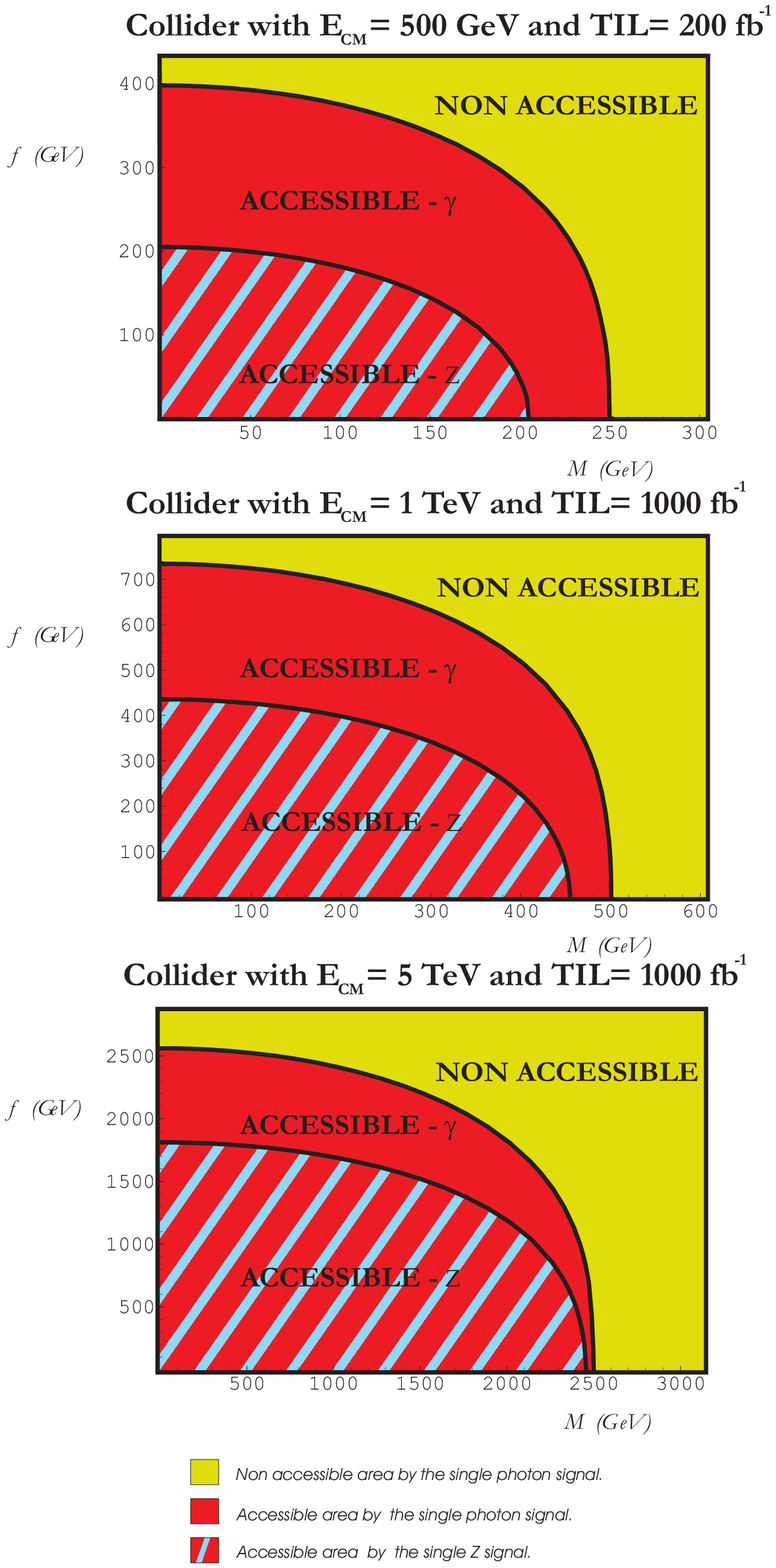}}
\vspace{-.5 cm}
\begin{figure}[h]
\caption{\footnotesize{Predicted experimental accessible 
regions in the $f-M$ 
plane via single-photon and single-Z processes. Different 
center-of-mass energies
($E_{CM}$) and total integrated luminosities (TIL) at 
future $e^+ e^-$ linear colliders are considered.}}
\end{figure}

\section{Summary and conclusions}

In this work a brane-world scenario
where the brane tension scale $f$ is much
smaller than the fundamental gravitational scale $M_F$ has been 
considered. 
For this case, the relevant low-energy degrees of freedom 
are the brane SM
particles and the brane oscillations or branons. From the 
corresponding
effective action, and including also the effects of a possible 
branon non-zero
mass, we have obtained the relevant Feynman rules for the 
couplings of branons 
to SM
particles. They have allowed us to compute the decay rates and
cross-sections for the different processes relevant for branon 
production in
electron-positron colliders. We have used the information 
coming from LEP in
order to get different exclusion plots on the branon mass and 
the tension scale plane. Single-photon production turns 
out to be the most efficient
process in order to set bounds on the brane parameters $f$ and $M$. 
We have
also extended the analysis to future electron-positron colliders. The
corresponding exclusion plots that could be obtained in  
case that branons were not observed, have been also shown. 

The work
presented here should be complemented with a parallel analysis for
electron-hadron colliders such as HERA and hadron-hadron colliders 
such as 
the Tevatron or the LHC, and with other bounds on $f$ and $M$ that 
could come
from astrophysics and cosmology. Work is already in progress in these 
directions. In particular, concerning cosmology, it
is interesting to notice that 
the allowed range of parameters suggests that branons could have
weak couplings and large masses. This makes them natural dark matter
candidates. In fact, an explicit calculation shows that 
their relic abundance can be cosmologically relevant and could
account for the fraction of one third of the total energy density of
the universe in the form of dark matter presently favoured by
observations. These results will be presented elsewhere 
\cite{CDM}.
  
 \vspace{.5cm}
 {\bf Acknowledgements:} This work
has been partially supported by the DGICYT (Spain) under the 
project numbers
 PB98-0782, AEN99-0305, FPA 2000-0956 and BFM2000-1326,
and also by the director, Office of Science,
Office of High Energy and Nuclear Physics 
of the U.S. Departament of
Energy under Contract DE-AC03-76SF00098.
A.D. acknowledges support from the
Universidad Complutense del Amo Program.    \\

\newpage

\appendix
\section{Branon vertices}

We show the Feynman rules with leaving momenta for massive branons, 
including the interaction
vertices between two branons and the SM particles. The dependence
on the momenta of the different particles is explicitly written.
We have used the notation in \cite{DoMa}. Different expressions in
the case of massless branons have been derived in 
\cite{CrSt}. In that reference the fermionic equations
of motion are imposed on the Feynman rules. In any case, as 
a consistency check, we have seen that our final results 
are the same in both cases. 

\subsection{$V1[p_1, p_2, p_3, p_4]$}

\begin{table}[h]
\begin{tabular}{c c c}
\begin{tabular}{c c c}
$\pi^\alpha(p_3)$& &$\pi^\beta(p_4)$\\
 &\epsfysize=3.0 cm\epsfxsize=3.0
cm\epsfbox{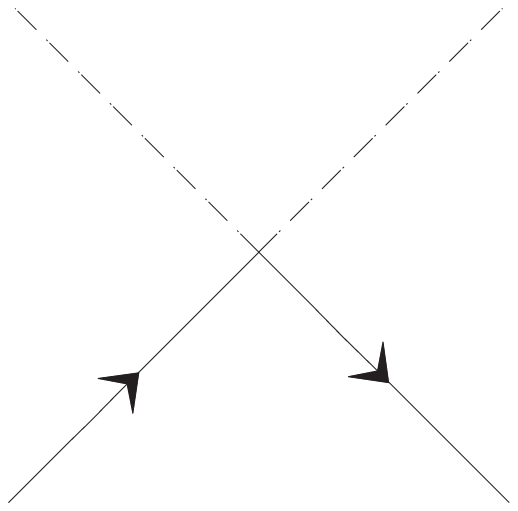}& \\
 $\bar\psi(p_1)$& &$\psi(p_2)$
\end{tabular}
 &
$\equiv$ & $V1[p_1, p_2, p_3, p_4]$
\end{tabular}
\end{table}
\begin{eqnarray}
V1&=& \frac{-i\delta^{\alpha\beta}}{4 f^4} \{\gamma^\mu
p_{4\mu}(p_3,p_1-p_2) +
\gamma^\mu p_{3\mu}(p_4,p_1-p_2)\nonumber\\
&-&\gamma^\mu
(p_{1\mu}-p_{2\mu})(\frac{3}{2}M^2+2(p_3,p_4))\nonumber\\
&+&4m_\psi((p_3,p_4)+M^2)\}.
\end{eqnarray}

\newpage
\subsection{$V2_{\mu\nu}^{ab}[p_1, p_2, p_3,p_4]$}

\begin{table}[h]
\begin{tabular}{c c c}
\begin{tabular}{c c c}
$\pi^\alpha(p_3)$& &$\pi^\beta(p_4)$\\
 &\epsfysize=3.0 cm\epsfxsize=3.0
cm\epsfbox{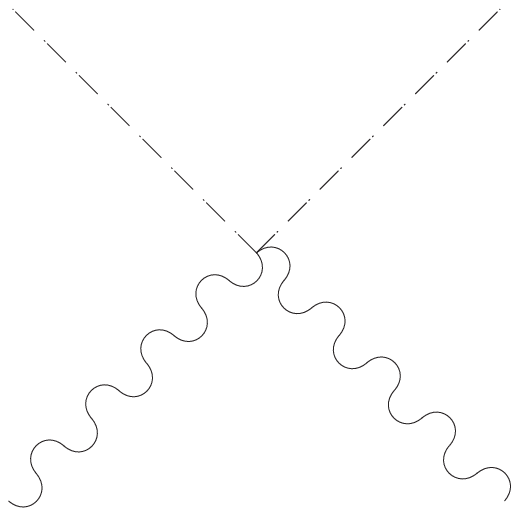}& \\
 $A^b_\nu(p_2)$& &$A^a_\mu(p_1)$
\end{tabular}
 &
$\equiv$ & $V2_{\mu\nu}^{ab}[p_1, p_2, p_3,p_4]$
\end{tabular}
\end{table}

\begin{eqnarray}
V2_{\mu\nu}^{ab}&=&
\frac{i\delta^{ab}\delta^{\alpha\beta}}{f^4}\{p_{1\nu}p_{3\mu}(p_2,p_4)+p_{3\nu}p_{2\mu}(p_1,p_4)
\nonumber\\
&+&p_{1\nu}p_{4\mu}(p_2,p_3)+p_{4\nu}p_{2\mu}(p_1,p_3)\nonumber\\
&-&\eta_{\mu\nu}((p_1,p_4)(p_2,p_3)+(p_1,p_3)(p_2,p_4)-(p_1,p_2)(p_3,p_4))\nonumber\\
&-&(p_1,p_2)(p_{4\nu}p_{3\mu}+p_{3\nu}p_{4\mu})-(p_3,p_4)(p_{1\nu}p_{2\mu})\nonumber\\
&-&\frac{1}{2}M_a^2(2p_{4\nu}p_{3\mu}+2p_{3\nu}p_{4\mu}
-2\eta_{\mu\nu}(p_3,p_4)-\eta_{\mu\nu}M^2)\},\nonumber\\
%
%
%
\end{eqnarray}
where we have used the flat background metric
to contract indices in the Fadeev-Popov Lagrangian.

\newpage
\subsection{$V3^a_\mu[p_3,p_4]$}
\begin{table}[h]
\begin{tabular}{c c c}
\begin{tabular}{c c c}
\hspace*{0cm}\vspace*{2cm}$\pi^\alpha(p_3)$& &\hspace*{-2cm}$\pi^\beta(p_4)$\\
\hspace*{-2cm}\vspace*{-2.5cm}$\bar\psi(p_1)$& &$\psi(p_2)$\\
&\hspace*{-1.2cm}\epsfysize=3.0
cm\epsfxsize=3.0cm\epsfbox{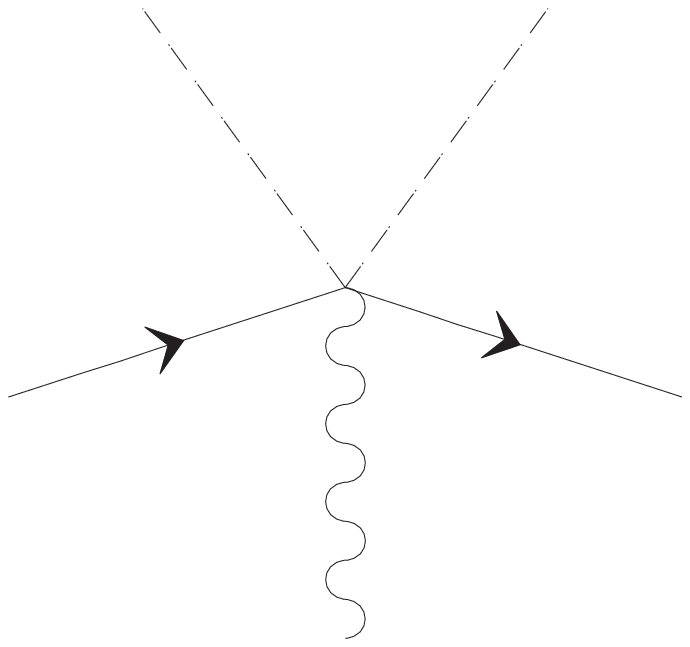}&
 \\
&\hspace*{-1.2cm}$A^a_\mu(p_5)$&
\end{tabular}
 &
$\equiv$ & $V3^a_\mu[p_3,p_4]$
\end{tabular}
\end{table}

\begin{eqnarray}
V3^a_\mu&=& \frac{-hT^a\delta^{\alpha\beta}}{4 f^4} \{2\gamma^\nu
p_{4\nu}p_{3\mu} +
2\gamma^\nu p_{3\nu}p_{4\mu}\nonumber\\
&+&\gamma_\mu(-3M^2 - 4 (p_3, p_4)\}(c_V-c_A\gamma_5).
\end{eqnarray}

\subsection{$V4_{\mu\nu\lambda}^{abc}[p_1,p_2,p_3,p_4,p_5]$}

\begin{table}[h]
\begin{tabular}{c c c}
\begin{tabular}{c c c}
\hspace*{0cm}\vspace*{2cm}$\pi^\alpha(p_3)$& &\hspace*{-2cm}$\pi^\beta(p_4)$\\
\hspace*{-2cm}\vspace*{-2.5cm}$A^b_\mu(p_1)$& &$A^c_\nu(p_2)$\\
&\hspace*{-1.2cm}\epsfysize=3.0 cm\epsfxsize=3.0
cm\epsfbox{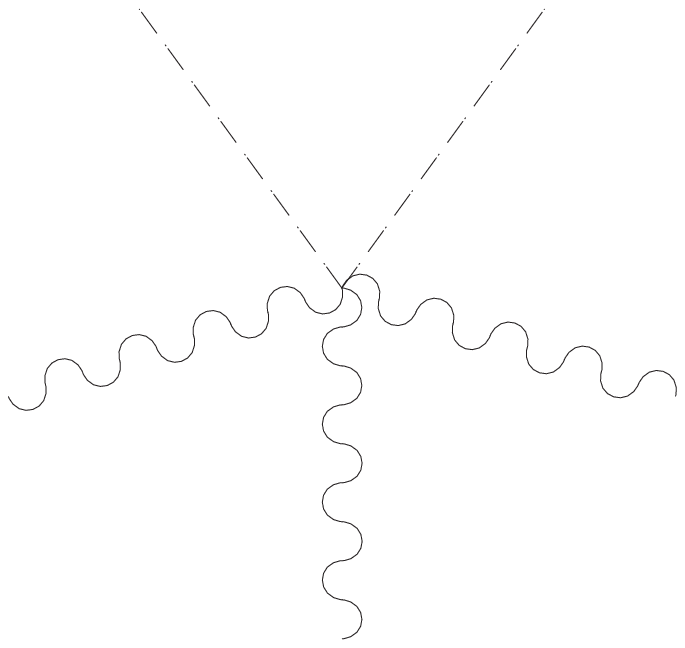}&
 \\
&\hspace*{-1.2cm}$A^a_\lambda(p_5)$&
\end{tabular}
 &
$\equiv$ & $V4_{\mu\nu\lambda}^{abc}[p_1,p_2,p_3,p_4,p_5]$
\end{tabular}
\end{table}

\begin{eqnarray}
 V4_{\mu\nu\lambda}^{abc}&=&\frac{h
C^{abc}\delta^{\alpha\beta}}{f^4}\{p_{1\nu}(p_{3\mu}p_{4\lambda}+p_{3\lambda}p_{4\mu})-p_{1\lambda}(p_{3\mu}p_{4\nu}+p_{3\nu}p_{4\mu})\\
&+&p_{2\lambda}(p_{3\nu}p_{4\mu}+p_{3\mu}p_{4\nu})-p_{2\mu}(p_{3\nu}p_{4\lambda}+p_{3\lambda}p_{4\nu})\nonumber\\
&+&p_{5\mu}(p_{3\nu}p_{4\lambda}+p_{3\lambda}p_{4\nu})-p_{5\nu}(p_{3\mu}p_{4\lambda}+p_{3\lambda}p_{4\mu})\nonumber\\
&+&\eta_{\lambda\nu}((p_3,p_4)(p_{2\mu}-p_{5\mu})+p_{4\mu}(p_5-p_2,p_3)+p_{3\mu}(p_5-p_2,p_4))\nonumber\\
&+&\eta_{\lambda\mu}((p_3,p_4)(p_{5\nu}-p_{1\nu})+p_{4\nu}(p_1-p_5,p_3)+p_{3\nu}(p_1-p_5,p_4))\nonumber\\
&+&\eta_{\mu\nu}((p_3,p_4)(p_{1\lambda}-p_{2\lambda})+p_{4\lambda}(p_2-p_1,p_3)+p_{3\lambda}(p_2-p_1,p_3))\}\nonumber
\end{eqnarray}

%
%
%




\newpage

\subsection{$V5[p_1, p_2, p_3, p_4]$}

\begin{table}[h]
\begin{tabular}{c c c}
\begin{tabular}{c c c}
$\pi^\alpha(p_3)$& &$\pi^\beta(p_4)$\\
 &\epsfysize=3.0 cm\epsfxsize=3.0
cm\epsfbox{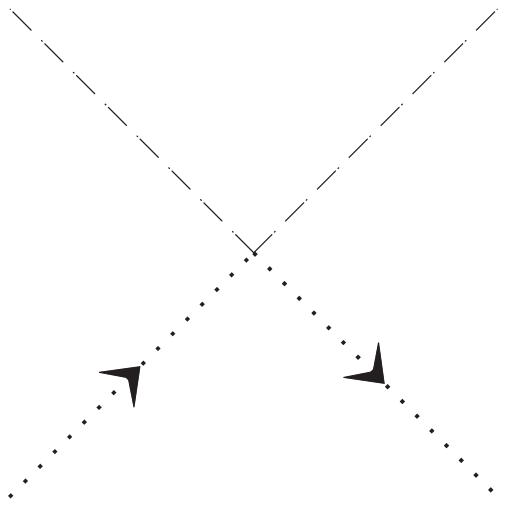}& \\
 $\phi^\dagger(p_1)$& &$\phi(p_2)$
\end{tabular}
 &
$\equiv$ & $V5[p_1, p_2, p_3, p_4]$
\end{tabular}
\end{table}
\begin{eqnarray}
V5&=& \frac{i\delta^{\alpha\beta}}{f^4} \left\{-[(p_3,p_4)+M^2]
[(p_1,p_2)+m_\phi^2]\right.\nonumber\\
&+&\left.(p_1,p_3)(p_4,p_2)+(p_2,p_3)(p_4,p_1)
+\frac{1}{2}M^2(p_1,p_2)\right\}.
\end{eqnarray}


\newpage
\subsection{$V6^a_\mu[p_3,p_4]$}
\begin{table}[h]
\begin{tabular}{c c c}
\begin{tabular}{c c c}
\hspace*{0cm}\vspace*{2cm}$\pi^\alpha(p_3)$& &\hspace*{-2cm}$\pi^\beta(p_4)$\\
\hspace*{-2cm}\vspace*{-2.5cm}$\phi^\dagger(p_1)$& &$\phi(p_2)$\\
&\hspace*{-1.2cm}\epsfysize=3.0
cm\epsfxsize=3.0cm\epsfbox{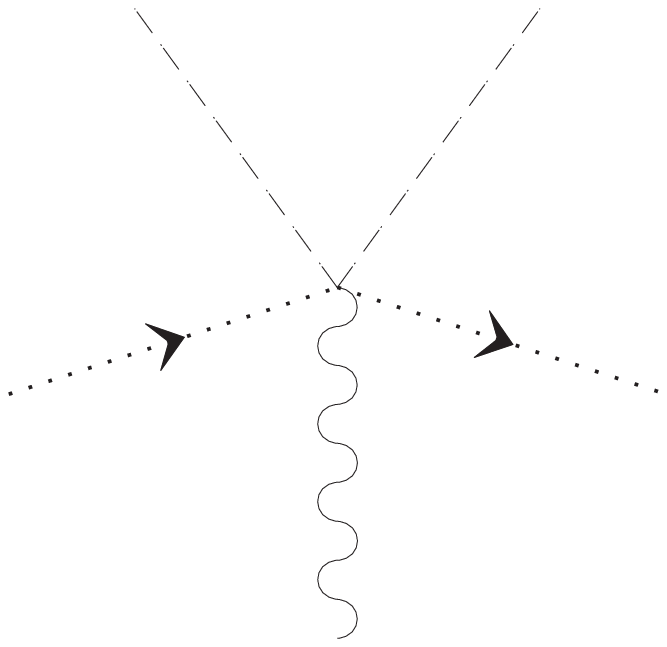}&
 \\
&\hspace*{-1.2cm}$A^a_\mu(p_5)$&
\end{tabular}
 &
$\equiv$ & $V6^a_\mu[p_3,p_4]$
\end{tabular}
\end{table}

\begin{eqnarray}
V6^a_\mu&=& \frac{hT^a\delta^{\alpha\beta}}{f^4} \{(p_{1}-p_{2})_\mu[(p_3,
p_4) +
\frac{1}{2}M^2]\nonumber\\
&+&(p_1-p_2,p_3)p_{4\mu}+(p_1-p_2,p_4)p_{3\mu}\}.
%
%
\end{eqnarray}


%
\subsection{$V7^{ab}_{\mu\nu}[p_3,p_4]$}
\begin{table}[h]
\begin{tabular}{c c c}
\begin{tabular}{c c c}
\hspace*{0cm}\vspace*{1.2cm}$\pi^\alpha(p_3)$& &\hspace*{-2cm}$\pi^\beta(p_4)$\\
\hspace*{-2cm}\vspace*{-1.7cm}$\phi^\dagger(p_1)$& &$\phi(p_2)$\\
&\hspace*{-1.2cm}\epsfysize=3.0
cm\epsfxsize=3.0cm\epsfbox{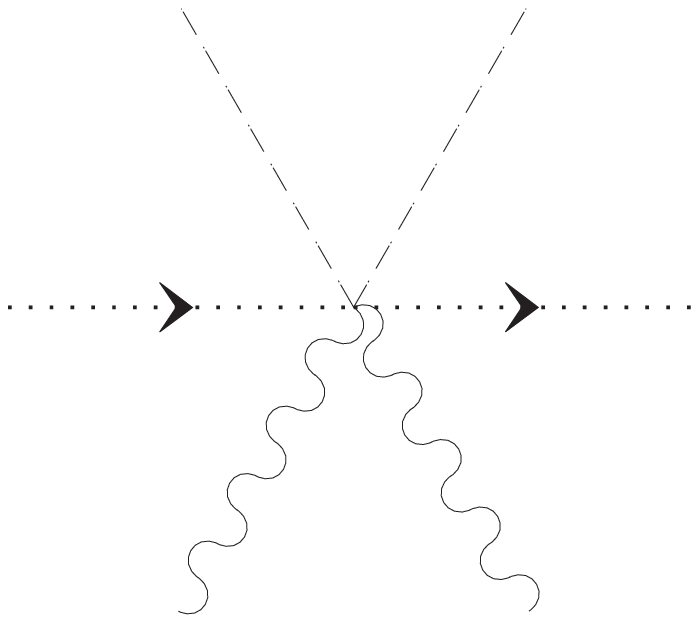}&
 \\
\hspace*{0cm}$A^a_\mu(p_5)$& &\hspace*{-2cm}$A^b_\nu(p_6)$
\end{tabular}
 &
$\equiv$ & $V7^{ab}_{\mu\nu}[p_3,p_4]$
\end{tabular}
\end{table}

\begin{eqnarray}
V7^{ab}_{\mu\nu}&=& \frac{-ih^2\{T^a,T^b\}\delta^{\alpha\beta}}{f^4} \{[(p_3, p_4)
+ \frac{1}{2}M^2]\eta_{\mu\nu}
%
-p_{3\mu} p_{4\nu}-p_{4\mu} p_{3\nu}\}.
\end{eqnarray}
\newpage
\subsection{$V8^{abcd}_{\mu\nu\rho\sigma}[p_3,p_4]$}
\begin{table}[h]
\begin{tabular}{c c c}
\begin{tabular}{c c c}
\hspace*{0cm}\vspace*{1.2cm}$\pi^\alpha(p_3)$& &\hspace*{-2cm}$\pi^\beta(p_4)$\\
\hspace*{-2cm}\vspace*{-1.7cm}$A^c_\rho(p_5)$& &$A^d_\sigma(p_6)$\\
&\hspace*{-1.2cm}\epsfysize=3.0
cm\epsfxsize=3.0cm\epsfbox{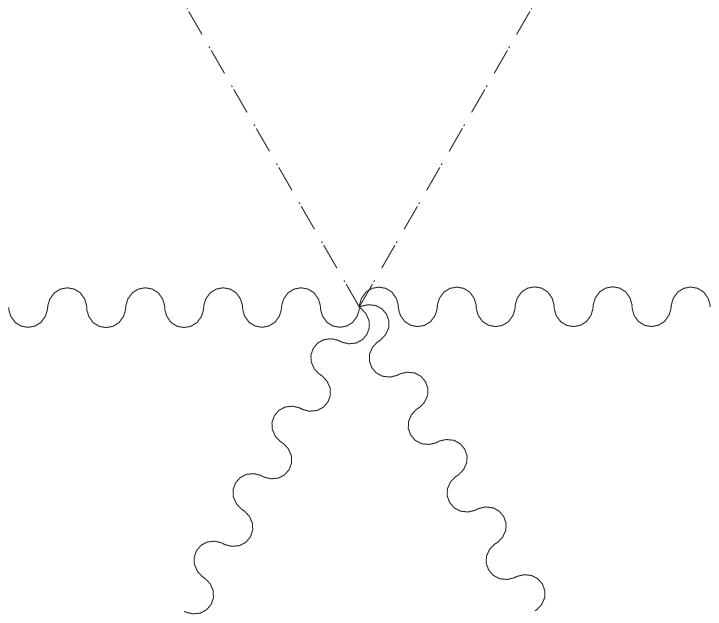}&
 \\
\hspace*{0cm}$A^a_\mu(p_1)$& &\hspace*{-2cm}$A^b_\nu(p_2)$
\end{tabular}
 &
$\equiv$ & $V8^{abcd}_{\mu\nu\rho\sigma}[p_3,p_4]$
\end{tabular}
\end{table}

\begin{eqnarray}
V8^{abcd}_{\mu\nu\rho\sigma}[p_3,p_4]&=&
\frac{ih^2\delta^{\alpha\beta}}{f^4}
\{C^{eab}C^{ecd}[\eta_{\nu\sigma}(p_{3\rho}p_{4\mu}+p_{3\mu}p_{4\rho})\nonumber\\
&-&
\eta_{\nu\rho}(p_{3\sigma}p_{4\mu}+p_{3\mu}p_{4\sigma})\nonumber\\
&+&
\eta_{\mu\sigma}(\eta_{\nu\rho}(p_3,p_4)-p_{3\rho}p_{4\nu}-p_{3\nu}p_{4\rho})\nonumber\\
&-&
\eta_{\mu\rho}(\eta_{\nu\sigma}(p_3,p_4)-p_{3\sigma}p_{4\nu}-p_{3\nu}p_{4\sigma})]\nonumber\\
&+&
C^{eac}C^{ebd}[\eta_{\nu\mu}(p_{3\rho}p_{4\sigma}+p_{3\sigma}p_{4\rho})\nonumber\\
&-&
\eta_{\nu\rho}(p_{3\sigma}p_{4\mu}+p_{3\mu}p_{4\sigma})\nonumber\\
&+&
\eta_{\mu\sigma}(\eta_{\nu\rho}(p_3,p_4)-p_{3\rho}p_{4\nu}-p_{3\nu}p_{4\rho})\nonumber\\
&-&
\eta_{\sigma\rho}(\eta_{\nu\mu}(p_3,p_4)-p_{3\mu}p_{4\nu}-p_{3\nu}p_{4\mu})]\nonumber\\
&+&
C^{ead}C^{ebc}[\eta_{\nu\mu}(p_{3\rho}p_{4\sigma}+p_{3\sigma}p_{4\rho})\nonumber\\
&-&
\eta_{\nu\sigma}(p_{3\rho}p_{4\mu}+p_{3\mu}p_{4\rho})\nonumber\\
&+&
\eta_{\mu\rho}(\eta_{\nu\sigma}(p_3,p_4)-p_{3\sigma}p_{4\nu}-p_{3\nu}p_{4\sigma})\nonumber\\
&-&
\eta_{\sigma\rho}(\eta_{\nu\mu}(p_3,p_4)-p_{3\mu}p_{4\nu}-p_{3\nu}p_{4\mu})]\}.
%
%
\end{eqnarray}

\newpage

\section{Probability amplitudes}

The probability amplitudes are easily  calculated from
the Feynman rules introduced before:

\subsection{$M1$: $\bar\psi(p_1),\psi(p_2)
\longrightarrow\pi(k_1)\pi(k_2)A^a_\mu(q)$}

There are four different diagrams that contribute to the 
tree-level amplitude of this process:

\begin{figure}[h]
\epsfxsize=13.0 cm\epsfbox{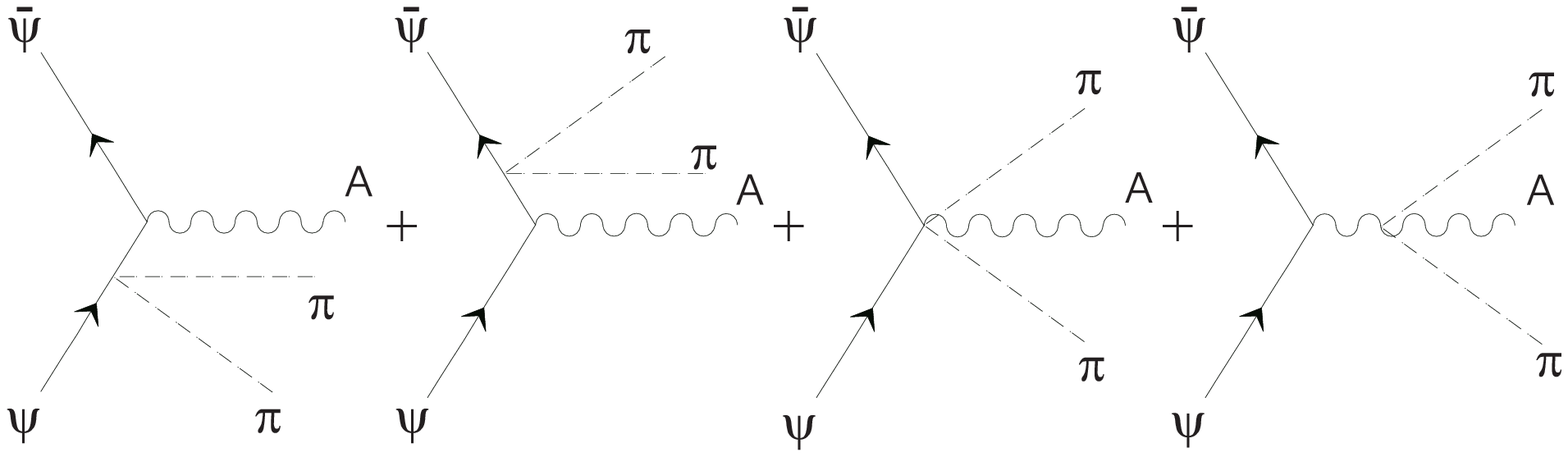}
\end{figure}
which is given by the following expression:
\begin{eqnarray}
&&M1[p_1,p_2,k_1,k_2,q]= \bar
v[p1]\{V3^a_\mu[k_1,k_2]\varepsilon^{\mu,a}[q,\sigma1]\label{Z0eamp}\\
&&+V11^a_\mu\varepsilon^{\mu,a}[q,\sigma1]i\Delta_{(f)}[p_1-q]V1[-(p_1-q),-p_2,
k_1, k_2]\nonumber\\
&&+V1[-p_1,-(p_2-q), k_1,
k_2]i\Delta_{(f)}[-(p_2-q)]V11^a_\mu\varepsilon^{\mu,a}[q,\sigma1]\nonumber\\
&&+V11^c_\lambda
i\Delta^{\lambda\nu,cb}_{(A)}[(p_1+p_2)]V2_{\mu\nu}^{ab}[q,
-(p_1+p_2), k_1,k_2]\varepsilon^{\mu,a}[q,\sigma1]\}u[p2].\nonumber
\end{eqnarray}
The fermion $(i\Delta_{(f)})$ and gauge boson $(i\Delta_{(A)})$
propagators  have the usual
expression in the Standard Model in renormalizable gauges,
and we have used the standard vertex between 
gauge and fermions fields:

\begin{table}[h]
\begin{tabular}{c c c}
\begin{tabular}{c c c}
&$A^a_\mu(p_3)$&\\
 &\epsfysize=3.0 cm\epsfxsize=3.0
cm\epsfbox{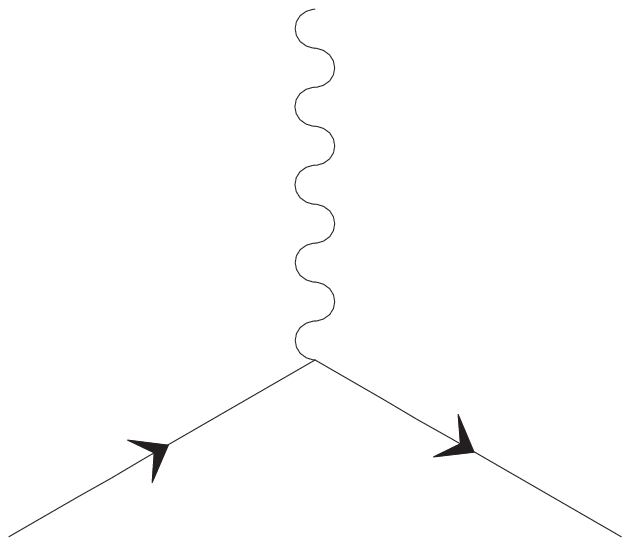}& \\ $\bar\psi(p_1)$& &$\psi(p_2)$
\end{tabular}
 &
$\equiv$ & $V11^a_\mu \equiv 
h\gamma_\mu T^a(c_V-c_A\gamma_5)$
\end{tabular}
\end{table}
%
%

\subsection{$M2$: $A^a_\mu(q)\longrightarrow\pi(k_1)\pi(k_2)\bar\psi(p_1),\psi(p_2)$}
If there are massive gauge bosons in the theory, branons 
contribute to their decay
through these processes, whose probability 
amplitude can be related to the 
previous one by: 
\begin{eqnarray}
&&M2[p_1,p_2,k_1,k_2,q]=M1[-p_1,-p_2,k_1,k_2,-q], \label{Z0amp}
\end{eqnarray}
where in
the right hand side, the spinors have been changed as follows 
$\bar v[p_1]\rightarrow \bar u[p_1]$ and 
$u[p_2]\rightarrow v[p_2]$.


\section{Integrals over two-body phase space}

Integrals over the two-body phase space  of branons
can be easily performed using:

\begin{eqnarray}
I_{\mu\nu\rho\sigma}&\equiv& \int d^4k_1d^4k_2
\delta(k_1^2-M^2)\delta(k_2^2-M^2)\delta^{(4)}(k-k_1-k_2)k_{1\mu}k_{1\nu}k_{2\rho}k_{2\sigma}
)\nonumber\\
&=&C_1k_{\mu}k_{\nu}k_{\rho}k_{\sigma}
+C_2k^4(\eta_{\mu\nu}\eta_{\rho\sigma}
+\eta_{\mu\rho}\eta_{\nu\sigma}+\eta_{\mu\sigma}\eta_{\nu\rho})\\
&+&C_3k^2(k_{\mu}k_{\nu}\eta_{\rho\sigma}
+k_{\rho}k_{\sigma}\eta_{\mu\nu})\nonumber \\
&+&C_4k^2(k_{\mu}k_{\rho}\eta_{\nu\sigma}
+k_{\nu}k_{\sigma}\eta_{\mu\rho}+
k_{\mu}k_{\sigma}\eta_{\rho\nu}
+k_{\rho}k_{\nu}\eta_{\mu\sigma})\nonumber
\end{eqnarray}

\begin{eqnarray}
I^m_{\mu\nu}&\equiv&\int d^4k_1d^4k_2
\delta(k_1^2-M^2)\delta(k_2^2-M^2)\delta^{(4)}(k-k_1-k_2)k_{1\mu}k_{2\nu}\nonumber\\
&=&C^m_1k_{\mu}k_{\nu}+C^m_2k^2\eta_{\mu\nu}
\end{eqnarray}

\begin{eqnarray}
I^s_{\mu\nu}&\equiv&\int d^4k_1d^4k_2
\delta(k_1^2-M^2)\delta(k_2^2-M^2)\delta^{(4)}(k-k_1-k_2)k_{1\mu}k_{1\nu}\nonumber\\
&=&C^s_1k_{\mu}k_{\nu}+C^s_2k^2\eta_{\mu\nu}
\end{eqnarray}

And the same result by changing $k_1$ for $k_2$.

\begin{eqnarray}
I^0&\equiv&\int d^4k_1d^4k_2
\delta(k_1^2-M^2)\delta(k_2^2-M^2)\delta^{(4)}(k-k_1-k_2)\nonumber\\
&=&C_0
\end{eqnarray}
where

\begin{eqnarray}
C_0&=& \frac{\pi}{2}\sqrt{1-\frac{4M^2}{k^2}}\nonumber\\
C_1&=&
\frac{\pi}{60k^4}\sqrt{1-\frac{4M^2}{k^2}}(k^4+2k^2M^2+6M^4)\nonumber\\
C_1^m&=&
\frac{\pi}{12k^2}\sqrt{1-\frac{4M^2}{k^2}}(k^2+2M^2)\nonumber\\
C_1^s&=&
\frac{\pi}{6k^2}\sqrt{1-\frac{4M^2}{k^2}}(k^2-M^2)\nonumber\\
C_2&=&
\frac{\pi}{480k^4}\sqrt{1-\frac{4M^2}{k^2}}(k^2-4M^2)^2\nonumber\\
C_2^m&=&
\frac{\pi}{24k^2}\sqrt{1-\frac{4M^2}{k^2}}(k^2-4M^2)\nonumber\\
C_2^s&=&-
\frac{\pi}{24k^2}\sqrt{1-\frac{4M^2}{k^2}}(k^2-4M^2)\nonumber\\
C_3&=&-
\frac{\pi}{240k^4}\sqrt{1-\frac{4M^2}{k^2}}(3k^4-14k^2M^2+8M^4)\nonumber\\
C_4&=&
\frac{\pi}{120k^4}\sqrt{1-\frac{4M^2}{k^2}}(k^4-3k^2M^2-4M^4)
\end{eqnarray}

\section{Polarized cross sections}
In this section we give the expressions for the cross sections
of the process:
\begin{eqnarray}
\sigma_A:
\bar\psi(p_1),\psi(p_2)\longrightarrow\pi(k_1),\pi(k_2),A^a_\mu(q)
\end{eqnarray}
for the different polarizations of the outgoing gauge field. We 
have  used the definition:
\begin{eqnarray}
q^\mu&=&(q_0,
|\overrightarrow{q}|\sin\theta\cos\phi,
|\overrightarrow{q}|\sin\theta\sin\phi,
|\overrightarrow{q}|\cos\theta),
\end{eqnarray}
The polarization vectors read:
\begin{eqnarray}
\epsilon^\mu_0&=&\frac{1}{M}(|\overrightarrow{q}|,
q_0\sin\theta\cos\phi,
q_0\sin\theta\sin\phi,
q_0\cos\theta);\\
\epsilon^\mu_\pm&=&\frac{1}{\sqrt{2}}(0,
-\cos\theta\cos\phi\pm i \sin\phi,
-\cos\theta\sin\phi\mp i \cos\phi,
\sin\theta).\nonumber
\end{eqnarray}
For the longitudinal polarization ($\epsilon^\mu_0$), we get:
\begin{eqnarray}
\frac{d\sigma_A}{dxd\cos\theta}&=&
\frac{\vert h\vert^2}{4\pi} \frac{(c_V^2+c_A^2)M_a^2 n
}{122880f^8\pi^2(s-M_a^2)^2}
\sqrt{1-\frac{4M^2}{s(1-x)+M_a^2}}\nonumber\\
%
&&\sqrt{s(sx^2-4M_a^2)}\{5M^2(2(s(1-x)
+M_a^2)-5M^2)\{sx^2\sin^2\theta\}\nonumber\\
%
%
&+&\frac{(s(1-x)+M_a^2-4M^2)^2}{(4M_a^2(M_a^2+s(1-x))
+s(sx^2-4M_a^2)\sin^2\theta)^2}\nonumber\\
&&\{32(s(1-x)+M_a^2)^3(M_a^2+s)^2-16(s(1-x)+M_a^2)^2\nonumber\\
&&[M_a^6+2M_a^4sx(1-x)+M_a^2s^2(9-2x^2)\nonumber \\
&+&2s^3(1-x-x^2)]
\sin^2\theta-4s[4M_a^8+M_a^6s(12-20x+15x^2)\nonumber\\
&-&M_a^4s^2(4+8x-37x^2+11x^3+4x^4)\nonumber\\
&+&M_a^2s^3(-12+28x-7x^2-6x^3-10x^4+4x^5)\nonumber\\
&+&s^4x^2(3-7x+2x^2+2x^3)]\sin^4\theta\nonumber \\
&+&2s^3(2+x^2)(sx^2-4M_a^2)^2\sin^6\theta\}\}.
\label{long}
\end{eqnarray}
The contributions of the transverse polarizations ($\epsilon_\mu^+$
or $\epsilon_\mu^-$) are the same and they are given by:
\begin{eqnarray}
\frac{d\sigma_A}{dxd\cos\theta}&=&
\frac{\vert h\vert^2}{4\pi}\frac{(c_V^2+c_A^2)n}{122880f^8\pi^2(s-M_a^2)^2}
\sqrt{1-\frac{4M^2}{s(1-x)+M_a^2}}\nonumber\\
%
&&\sqrt{s(sx^2-4M_a^2)}\{10M^2 M_a^4(2(s(1-x)+M_a^2)-5M^2)
\{2-\sin^2\theta\}\nonumber\\
%
%
&+&\frac{(s(1-x)+M_a^2-4M^2)^2}{(4M_a^2(M_a^2+s(1-x))
+s(sx^2-4M_a^2)\sin^2\theta)^2}\nonumber\\
&&\{16M_a^4(s(1-x)+M_a^2)^2[9M_a^4+M_a^2s(6-4x)+s^2(5-4x+2x^2)]
\nonumber\\
&+&2[-36M_a^{12}+2M_a^{10}s(-108+40x+x^2)\nonumber \\
&-&4M_a^8s^2(68-72x+6x^2+x^3)\nonumber\\
&+&M_a^6s^3(-136+200x-54x^2-26x^3+3x^4)\nonumber \\
&-&M_a^4s^4(44-80x+34x^2+12x^3-13x^4+x^5)\nonumber\\
&+&M_a^2s^5x(-8+28x-34x^2+21x^3-6x^4)\nonumber\\
&-&s^6x^2(-2+4x-3x^2+x^3)]\sin^2\theta+s[120M_a^{10}\nonumber \\
&-&2M_a^8s(-148+76x+13x^2)-M_a^6s^2(-8+48x+62x^2\nonumber \\
&-&34x^3+x^4)
+M_a^4s^3(24+8x+10x^2+4x^3-3x^4+x^5)\nonumber\\
&+&M_a^2s^4x^2(-18+10x-11x^2+2x^3)+s^5x^4(3(1-x)\nonumber\\
&&+2x^2)]\sin^4\theta-s^2(sx^2-4M_a^2)^2[4M_a^4-2M_a^2s+s^2x^2]
\sin^6\theta\}\}
\label{trans}
\end{eqnarray}


\end{document}